\documentclass[12pt,prd, aps, showpacs, superscriptaddress,floatfix]{revtex4}

\usepackage{amsmath}
\usepackage{graphicx}
\usepackage{mathrsfs}
\usepackage{hyperref}
\usepackage{color}

\usepackage{appendix}

\newcommand{\be}{\begin{equation}}
\newcommand{\ee}{\end{equation}}
\newcommand{\bea}{\begin{eqnarray}}
\newcommand{\eea}{\end{eqnarray}}
\newcommand{\ket}[1]{\ensuremath{| {#1} \rangle }}
\newcommand{\bra}[1]{\ensuremath{\langle {#1} |}}
\newcommand{\Romatre}{Dipartimento di Fisica, Universit\`a  Roma Tre and INFN, Sezione di Roma Tre, Via della Vasca Navale 84, I-00146 Rome, Italy}
\newcommand{\Sissa}{SISSA, Via Bonomea 265, I-34136, Trieste,  and INFN Sezione di Roma La Sapienza Piazzale Aldo Moro 5, 00185 Roma, Italy}
\newcommand{\soton}{School of Physics and Astronomy, University of Southampton,  Southampton SO17 1BJ, UK}
\newcommand{\Romadue}{Dipartimento di Fisica and INFN, Universit\`a di Roma ``Tor Vergata", Via della Ricerca Scientifica 1, I-00133 Roma, Italy}
\newcommand{\LaSapienza}{Physics Department and INFN Sezione di Roma La Sapienza Piazzale Aldo Moro 5, 00185 Roma, Italy}
\newcommand{\cern}{PH-TH, CERN, CH-1211, Geneva 23, Switzerland}
\begin{document}

\title{QED Corrections to Hadronic Processes in Lattice QCD}

\author{N.Carrasco}\affiliation{\Romatre}
\author{V.Lubicz}\affiliation{\Romatre}
\author{G.Martinelli}\affiliation{\Sissa}
\author{C.T.Sachrajda}\affiliation{\soton}
\author{N.Tantalo}\affiliation{\cern}\affiliation{\Romadue}
\author{C.Tarantino}\affiliation{\Romatre}
\author{M.Testa}\affiliation{\LaSapienza}


\begin{abstract}
\vskip 0.2cm
In this paper, for the first time a method is proposed to compute electromagnetic effects in hadronic processes using lattice simulations.  The method can be applied, for example, to the leptonic and semileptonic decays of light or heavy pseudoscalar mesons. For these quantities the presence of infrared divergences in intermediate stages of the calculation makes the procedure much more complicated  than is the case for the hadronic spectrum, for which calculations  already exist. In order to compute the physical widths, diagrams with virtual photons must be combined with those corresponding to the emission of real photons. Only in this way do the infrared divergences cancel as first understood by Bloch and Nordsieck in 1937.  We present a detailed analysis of the method for the leptonic decays of a pseudoscalar meson. The implementation of our method, although challenging,  is within reach of the present lattice technology.  
\end{abstract}

\maketitle

\section{Introduction} Precision flavour physics is a particularly powerful tool for exploring the limits of the Standard Model (SM) of particle physics and in searching for inconsistencies which would signal the existence of new physics. An important component of this endeavour is the over-determination of the elements of the Cabibbo-Kobayashi-Maskawa (CKM) matrix from a wide range of weak processes. The precision in extracting CKM matrix elements is generally limited by our ability to quantify hadronic effects and the main goal of large-scale simulations using the lattice formulation of QCD is the  \emph{ab-initio} evaluation of the non-perturbative QCD effects in physical processes. 
The recent, very impressive, improvement in lattice computations has led to a precision approaching $O(1\%)$ for a number of quantities (see e.g. Ref.\,\cite{Aoki:2013ldr} and references therein) and therefore in order to make further progress electromagnetic effects (and other isospin-breaking contributions) have to be considered. The question of how to include electromagnetic effects in the hadron spectrum and in the determination of quark masses in ab-initio
lattice calculations was addressed for the first time in~\cite{Duncan:1996xy}. 
Much theoretical and algorithmic progress has been made following this pioneering work, particularly in recent years, leading to remarkably accurate determinations of the charged-neutral mass splittings of light pseudoscalar mesons and light baryons (see Refs.~\cite{Borsanyi:2014jba,Basak:2014vca,deDivitiis:2013xla,Ishikawa:2012ix,Aoki:2012st,Blum:2010ym} for recent papers on the subject and Refs.~\cite{Tantalo:2013maa,Portellilattice} for reviews of these results and a discussion of the different approaches used to perform QED+QCD lattice calculations of the spectrum).

In the computation of the hadron spectrum there is a very significant simplification in that there are no infrared divergences. In this paper we propose a strategy to include 
electromagnetic effects in processes for which infrared divergences are present but which cancel in the standard way between diagrams containing different numbers of real and virtual photons~\cite{Bloch:1937pw}.  The presence of  infrared divergences in intermediate steps of the calculation requires the development of new methods. 
Indeed, in order to cancel the infrared divergences and obtain results for physical quantities,
radiative corrections from virtual and real photons must be combined.  
We stress that it is not sufficient simply to add the electromagnetic interaction to the quark action because amplitudes with different numbers of real 
photons must be evaluated separately, before being combined in the inclusive rate for a given process. 
In this paper for the first time we introduce and discuss a strategy to compute electromagnetic radiative corrections to leptonic decays of pseudoscalar mesons which can then be used to determine the corresponding CKM matrix elements. Although we present the explicit discussion for this specific set of processes, the method is more general and can readily be extended to generic processes including, for example, to semileptonic decays. 

We now focus on the leptonic decay of the charged pseudoscalar meson $P^+$. Let $\Gamma_0$ be the partial width for the decay $P^+\to\ell^+\nu_\ell$ where the charged lepton $\ell$ is an electron or a muon (or possibly a $\tau$) and $\nu_\ell$ is the corresponding neutrino. 
The subscript $0$ indicates that there are no photons in the final state. 
In the absence of electromagnetism, the non-perturbative QCD effects are contained in a single number, the decay constant $f_P$, defined by 
\begin{equation}\label{eq:fPdef}
\langle 0\,|\,\bar{q}_1\gamma^\mu\gamma^5\,q_2\,|\,P^+(p)\rangle=ip^\mu f_P\,,
\end{equation}
where $P^+$ is composed of the valence quarks $\bar q_1$ and ${q}_2$, and the axial current in (\ref{eq:fPdef}) is composed of the corresponding quark fields. There have been very many lattice calculations of the decay constants $f_\pi$,\,$f_K$,\,$f_{D_{(s)}}$ and $f_{B_{(s)}}$~\cite{Aoki:2013ldr}, some of which  are approaching $O(1\%)$ precision.  
As noted above, in order to determine the corresponding CKM matrix elements at this level of precision isospin breaking effects, including electromagnetic corrections, must be considered.  It will become clear in the following, and has been stressed in \cite{Bijnens:1993ae,Gasser:2010wz}, that it is not possible to give a physical definition of the decay constant $f_P$ in the presence of  electromagnetism, because of the contributions from diagrams in which the photon is emitted by the hadron and absorbed by the charged lepton. Thus the physical width is not just given in terms of the matrix element of the axial current  and  can only be obtained  by a full calculation of the electromagnetic corrections at a given order. 

The calculation of  electromagnetic effects leads to an immediate difficulty: $\Gamma_0$ contains infrared divergences and by itself is therefore unphysical. The well-known solution to this problem is to include the contributions from real photons. We therefore define $\Gamma_1(\Delta E)$ to be the partial width for the decay $P^+\to\ell^+\nu_\ell\,\gamma$ where the energy of the photon in the rest frame of $P^+$ is integrated from $0$ to $\Delta E$. The sum $\Gamma_0+\Gamma_1(\Delta E)$ is free from infrared divergences (although, of course, it does depend on the energy cut-off $\Delta E$). We restrict the discussion to $O(\alpha)$ corrections, where $\alpha$ is the electromagnetic fine-structure constant, and hence only consider a single photon.

The previous paragraph reminds us that the determination of the CKM matrix elements $V_{q_1q_2}$ at $O(\alpha)$ (i.e.~at $O(1\%)$ or better) from leptonic decays requires the evaluation of amplitudes with a real photon. The main goal of this paper is to suggest how such a calculation might be performed with non-perturbative accuracy. There are a number of technicalities which will be explained in the following sections, but here we present  a general outline of the proposed method.  We start with the experimental observable $\Gamma(\Delta E)$, the partial width for $P^+\to\ell^+\nu_\ell(\gamma)$. The final state consists either of $\ell^+\nu_\ell$ or of $\ell^+\nu_\ell\gamma$ where the energy of the photon in the centre-of-mass frame is smaller than $\Delta E$:
\begin{equation}\label{eq:Gamma01}
\Gamma(\Delta E)=\Gamma_0+\Gamma_1(\Delta E)\,.
\end{equation}   

In principle at least, $\Gamma_1(\Delta E)$ can be evaluated in lattice simulations by computing the amplitudes for a range of photon momenta and using the results to perform the integral over phase space. Such calculations would be very challenging. Since the computations are necessarily performed in finite volumes the available momenta are discrete, so that it would be necessary to choose the volumes appropriately and compute several correlation functions.
We choose instead to make use of the fact that a very soft photon couples to a charged hadron as if to an elementary particle; it does not resolve the structure of the hadron. We therefore propose to choose $\Delta E$ to be sufficiently small that the pointlike approximation can be used to calculate $\Gamma_1(\Delta E)$ in perturbation theory, treating $P^+$ as an elementary particle. On the other hand, $\Delta E$ must be sufficiently large that $\Gamma(\Delta E)$ can be measured experimentally. We imagine setting $\Delta E=O(10\,\textrm{-}\,20\,\mathrm{MeV})$ which satisfies both requirements. From Refs.\,\cite{Ambrosino:2005fw,Ambrosino:2009aa} we learn that resolutions on the energy of the photon in the rest frame of the decaying particle of this order are experimentally accessible.
In Appendix~\ref{sec:sd} we present a discussion, based on phenomenological analyses, of the uncertainties induced by treating the meson as elementary as a function of $\Delta E$.

It is necessary to ensure that the cancellation of infrared divergences occurs with good numerical precision leading to an accurate result for $\Gamma(\Delta E)$. Since $\Gamma_0$ is to be calculated in a Monte-Carlo simulation and $\Gamma_1(\Delta E)$ in perturbation theory using the pointlike approximation, this requires an intermediate step. We propose to rewrite Eq.\,(\ref{eq:Gamma01}) in the form
\begin{equation}\label{eq:master}
\Gamma(\Delta E)=\lim_{V\to\infty}(\Gamma_0-\Gamma_0^{\mathrm{pt}})+
\lim_{V\to\infty}(\Gamma_0^{\mathrm{pt}}+\Gamma_1(\Delta E))\,,
\end{equation}
where $V$ is the volume of the lattice.
$\Gamma_0^\mathrm{pt}$ is an unphysical quantity; it is the perturbatively calculated amplitude at $O(\alpha)$ for the decay $P^+\to\ell^+\nu_\ell$ with the $P^+$ treated as an elementary particle. In $\Gamma_0^\mathrm{pt}$ the finite-volume sum over the momenta of the photon is performed over the full range. The contributions from small momenta to $\Gamma_0$ and $\Gamma_0^\mathrm{pt}$ are the same and thus the infrared divergences cancel in the first term on the right-hand side of Eq.\,(\ref{eq:master}).  
Moreover, the infrared divergences in $\Gamma_0$ and $\Gamma_0^\mathrm{pt}$ are both equal and opposite to that in $\Gamma_1(\Delta E)$. The infrared divergences therefore cancel separately in each of the two terms on the right-hand side of Eq.\,(\ref{eq:master}) and indeed we treat each of these terms separately.    $\Gamma_0^{\mathrm{pt}}+\Gamma_1(\Delta E)$ is calculated in perturbation theory directly in infinite volume. The QCD effects in $\Gamma_0$ are calculated stochastically in a lattice simulation and the virtual photon is included explicitly in the Feynman gauge. For each photon momentum this is combined with $\Gamma_0^\mathrm{pt}$ and the difference is summed over the momenta and then the infinite-volume limit is taken. This completes the sketch of the proposed method, and in the remainder of this paper we explain the many technical issues which must be addressed. 

It will be helpful in the following to define $\Delta\Gamma_0(L)$ in terms of the first term on the right-hand side of Eq.\,(\ref{eq:master}): 
\begin{equation}\label{eq:DeltaGamma0def}
\Delta\Gamma_0(L)=\Gamma_0(L)-\Gamma_0^\mathrm{pt}(L)\,,
\end{equation}
where we have made the dependence on the volume explicit, $V=L^3$ and $L$ is the length of the lattice in any spacial direction (for simplicity we assume that this length is the same in all three directions). In analogy to Eq.\,(\ref{eq:Gamma01}) we also define the perturbative quantity
\begin{equation}\label{eq:Gamm01pt}
\Gamma^\mathrm{pt}(\Delta E)\equiv\Gamma^\mathrm{pt}_0+\Gamma_1(\Delta E)\,.
\end{equation}

We note that, since the sum of all the terms in Eq.\,(\ref{eq:master})  is gauge invariant  as is the perturbative rate $\Gamma^\mathrm{pt}(\Delta E)$, the combination  $\Delta\Gamma_0(L)$ is also gauge invariant, although each of the two terms is not.

The plan of this paper is as follows. In the next section we discuss the effective weak Hamiltonian and its renormalisation in the presence of electromagnetism. The structure of the calculation and the correlation functions which need to be calculated are presented in Sec.\,\ref{sec:structure}. 
The evaluation of the second term on the right-hand side of Eq.\,(\ref{eq:master}), $\Gamma^\mathrm{pt}(\Delta E)$, directly in infinite volume, is theoretically straightforward and we perform this calculation in~Sec.\,\ref{sec:Gamma1}. Sec.\,\ref{sec:ir} contains a detailed discussion of the regularisation and cancellation of infrared divergences in a finite volume. We put all the elements of the calculation together in Sec.\,\ref{sec:concs}, where we present a summary and the prospects for the implementation of the method in numerical simulations. 
There are two appendices. In Appendix\,\ref{sec:Wilsonmatching} we discuss the matching of the bare lattice operators used in the calculation of correlation functions and those defined in the $W$-regularisation which is a natural scheme used in the definition of the Fermi constant $G_F$ in the presence of electromagnetism. Finally in Appendix~\ref{sec:sd} we present some phenomenological estimates of the uncertainties due to the use of the point-like approximation for $P^+$ in the decay $P^+\to\ell^+\nu\gamma$.

In the remainder of the paper, to be specific we choose  $P^+ =\pi^+$ but the discussion generalizes trivially to other  pseudoscalar mesons with the obvious changes of flavour labels. The method does not require $P^+$ to be a light psuedo-Goldstone Boson nor on the use of chiral perturbation theory.

\section{Matching the effective local four-quark operator(s) onto the standard model}\label{sec:uv}

\begin{figure}[t]
\includegraphics[width=0.7\hsize]{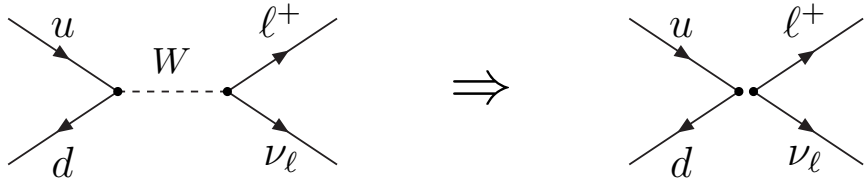}
\caption{Tree-level diagram for the process $u\bar{d}\to\ell^+\nu_\ell$ (left-hand diagram). In the effective theory the interaction is replaced by a local four-fermion operator (right-hand diagram).
\label{fig:DiagLO}}
\end{figure}

At lowest order in electromagnetic (and strong) perturbation theory the process $u\bar{d}\to\ell^+\nu_\ell$ proceeds by an $s$-channel $W$ exchange, see the left-hand diagram in Fig.\,\ref{fig:DiagLO}. Since the energy-momentum exchanges in this process are much smaller than $M_W$, it is standard practice to rewrite the amplitude in terms of a four-fermion local interaction:
\begin{equation}\label{eq:DeltaL0}
{\cal L}_W=-\frac{4G_F}{\sqrt{2}}\,V_{ud}^\ast\,\big(\bar{d}_L\gamma_\mu u_L\big)\,\big(\bar{\nu}_{\ell\,L}\gamma^\mu \ell_L\big)\,,
\end{equation}
where the subscript {\footnotesize $L$} represents \emph{left}, $\psi_L = \frac{(1-\gamma_5)}{2}\,  \psi$,  and $G_F$ is the Fermi constant. In performing lattice computations this replacement is necessary, since the lattice spacing $a$ is much greater than $1/M_W$, where $M_W$ is the mass of the $W$-Boson. When including the $O(\alpha)$ corrections, the ultra-violet contributions to the matrix element of the local operator are different to those in the Standard Model and in this section we discuss the matching factors which must be computed to determine the $O(\alpha)$ corrections to the $\pi^+\to\ell^+\nu_\ell$ decay from lattice computations of correlation functions containing the local operator in (\ref{eq:DeltaL0}).  Since the pion decay width is written in terms of $G_F$, it is necessary to start by revisiting the determination of the Fermi constant at $O(\alpha)$.

\subsection{Determination of the Fermi constant, $\mathbf{G_F}$}

$G_F$ is conventionally taken from the measured value of the muon lifetime using the expression~\cite{Berman:1958ti,Kinoshita:1958ru}
\begin{equation}\label{eq:muonlifetime}
\frac{1}{\tau_\mu}=\frac{G_F^2m_\mu^5}{192\pi^3}\left[1-\frac{8m_e^2}{m_\mu^2}\right]\left[1+\frac{\alpha}{2\pi}\left(\frac{25}{4}-\pi^2\right)\right],\end{equation}
leading to the value $G_F=1.16634\times 10^{-5}\,\mathrm{GeV}^{-2}$. (For an extension of Eq.\,(\ref{eq:muonlifetime}) to $O(\alpha^2)$ and the inclusion of higher powers of $\rho\equiv(m_e/m_\mu)^2$ see Sec.\,10.2 of~\cite{Beringer:1900zz}. The Particle Data Group~\cite{Beringer:1900zz} quote the corresponding value of the Fermi constant to be $G_F=1.1663787(6)\times 10^{-5}\,\mathrm{GeV}^{-2}$.)

\begin{figure}[t]
\begin{center}
\includegraphics[width=0.25\hsize]{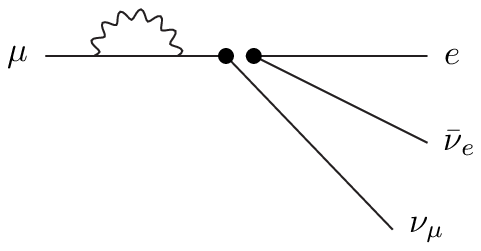}\qquad
\includegraphics[width=0.25\hsize]{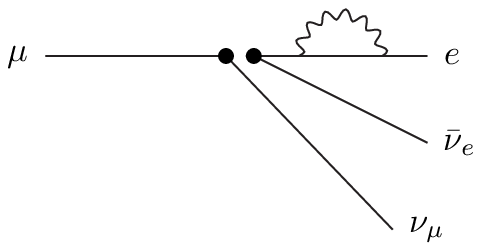}\qquad
\includegraphics[width=0.25\hsize]{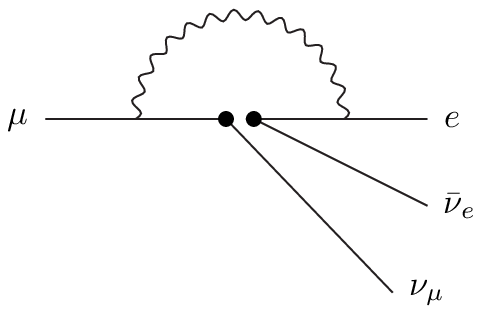}
\end{center}
\caption{Diagrams contributing to the $O(\alpha)$ corrections to muon decay; see Eq.\,(\ref{eq:muonlifetime}). The curly line represents the photon.\label{fig:muondecay}}
\end{figure}

Eq.\,(\ref{eq:muonlifetime}) can be viewed as the definition of $G_F$. When calculating the Standard Model corrections to the muon lifetime many of the contributions are absorbed into $G_F$ and the remaining terms on the right-hand side of (\ref{eq:muonlifetime}) come from the diagrams in Fig.\,\ref{fig:muondecay}. Specifically in these diagrams the factor $1/k^2$ in the Feynman-gauge photon propagator is replaced by $1/k^2\times M_W^2/(M_W^2-k^2)$, where $k$ is the momentum in the propagator; this is called the $W$-regularisation of ultra-violet divergences. These diagrams are evaluated in the effective theory with the local four-fermion operator $(\bar\nu_\mu\gamma^\mu(1-\gamma^5)\mu)\,(\bar e\gamma^\mu(1-\gamma^5)\nu_e)$; the two currents are represented by the filled black circles in Fig.\,\ref{fig:muondecay}.

\begin{figure}[t]
\begin{center}
\includegraphics[width=0.4\hsize]{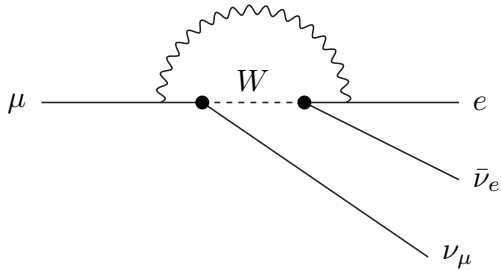}\qquad
\end{center}
\caption{Photon-$W$ box diagrams contributing to the $O(\alpha)$ corrections to muon decay in the Standard Model. The curly line represents the photon.\label{fig:gammaWbox}}
\end{figure}

An explanation of the reasoning behind the introduction of the W-regularisation is given in~\cite{Sirlin:1980nh}. The Feynman-gauge photon propagator is rewritten as two terms: 
\begin{equation}\label{eq:split}
\frac{1}{k^2}=\frac{1}{k^2-M_W^2}+\frac{M_W^2}{M_W^2-k^2}\,\frac1{k^2}
\end{equation}
and the ultra-violet divergent contributions come from the first term and are absorbed in the definition of $G_F$. In addition, the Standard-Model $\gamma$-$W$ box diagram in Fig.\,\ref{fig:gammaWbox} is ultra-violet convergent and is equal to the corresponding diagram in the effective theory (i.e. the third diagram in Fig.\,\ref{fig:muondecay}) with the W-regularisation, up to negligible corrections of $O(q^2/M_W^2)$, where $q$ is the four-momentum of the electron and its neutrino. Other electroweak corrections not explicitly mentioned above are all absorbed into $G_F$.

\subsection{$W$-regularisation and Weak Decays of Hadrons}

It is a particularly helpful feature that most of the terms which are absorbed into the definition of $G_F$ are common to other processes, including the leptonic decays of pseudoscalar mesons~\cite{Sirlin:1981ie,Braaten:1990ef}. 
There are however, some short-distance contributions which do depend on the electric charges of the individual fields in the four-fermion operators and these lead to a correction factor of $(1+\frac{2\alpha}{\pi}\log\frac{M_Z}{M_W})$ to $\Gamma_0$~\cite{Sirlin:1981ie}. This is a tiny correction ($\simeq 0.06\%$), but one which nevertheless can readily be included explicitly. 

The conclusion of the above discussion is that the evaluation of the amplitude for the process $\pi^+\to\ell^+\nu$ up to $O(\alpha)$ can be performed in the effective theory with the effective Hamiltonian
\begin{equation}\label{eq:Heff}
H_\mathrm{eff}=\frac{G_F}{\sqrt{2}}\,V_{ud}^\ast\left(1+\frac{\alpha}{\pi}\log\frac{M_Z}{M_W}\right)(\bar{d}\gamma^\mu (1-\gamma^5)u)\,
(\bar{\nu}_\ell\gamma_\mu(1-\gamma^5)\ell)\,,
\end{equation}
and with the Feynman-gauge photon propagator in the W-regularisation. The value of $G_F$ is obtained from the muon lifetime as discussed around Eq.\,(\ref{eq:muonlifetime}).

Of course we are not able to implement the W-regularisation directly in present day lattice simulations in which the inverse lattice spacing is much smaller than $M_W$. The relation between the operator in eq.\,(\ref{eq:Heff}) in the lattice and W regularisations can be computed in perturbation theory. Thus for example, with the Wilson action for both the gluons and fermions:
\begin{eqnarray}
O_1^\mathrm{W-reg}&=&\left(1+\frac{\alpha}{4\pi}\left(2\log a^2M_W^2-15.539\right)\right)O_1^\mathrm{bare} + \frac{\alpha}{4\pi}\ \left(
0.536\,O_2^\mathrm{bare} \right. \nonumber\\  
&&  \left.   \hspace{0.3in}+1.607\,O_3^\mathrm{bare}-3.214\,O_4^\mathrm{bare}-0.804\,O_5^\mathrm{bare} \right) \,,\label{eq:matchingWilson}
\end{eqnarray}
where 
\begin{align}
O_1&= (\bar{d}\gamma^\mu (1-\gamma^5)u)\,(\bar{\nu}_\ell\gamma_\mu(1-\gamma^5)\ell)&O_2&=
(\bar{d}\gamma^\mu (1+\gamma^5)u)\,(\bar{\nu}_\ell\gamma_\mu(1-\gamma^5)\ell)
\nonumber\\ 
O_3&= (\bar{d}(1-\gamma^5)u)\,(\bar{\nu}_\ell(1+\gamma^5)\ell)&O_4&= (\bar{d}(1+\gamma^5)u)\,(\bar{\nu}_\ell(1+\gamma^5)\ell)\label{eq:5ops}\\ 
O_5&=(\bar{d}\sigma^{\mu\nu}(1+\gamma^5) u)\,(\bar{\nu}_\ell\sigma_{\mu\nu}(1+\gamma^5)\ell)\,.\nonumber
\end{align}
The superscript ``bare" indicates that these are bare operators in the lattice theory and the presence of 5 operators on the right-hand side of Eq.\,(\ref{eq:matchingWilson}) is a consequence of the breaking of chiral symmetry in the Wilson theory. Using lattice actions with good chiral symmetry, such as domain wall fermions with a sufficiently large fifth dimension, only $O_1^\mathrm{bare}$ would appear on the right-hand side of Eq.(\ref{eq:matchingWilson}). The coefficients multiplying the operators depend of course on the lattice action being used. 
More details of the derivation of Eq.\,(\ref{eq:matchingWilson}) are presented in Appendix\,\ref{sec:Wilsonmatching}. 
Eq.\,(\ref{eq:matchingWilson}) is valid up to corrections of $O(\alpha_s(a)\,\alpha)$\,.

Having formulated the problem of calculating $\Gamma_0$ in terms of the evaluation of correlation functions involving the effective Hamiltonian in Eq.\,(\ref{eq:Heff}) we are now in a position to discuss the calculation of $\Delta\Gamma_0(L)$, the first term on the right-hand side of the master formula Eq.\,(\ref{eq:master}).

\section{Structure of the calculation}\label{sec:structure}

\begin{figure}[t]
\includegraphics[width=0.4\hsize]{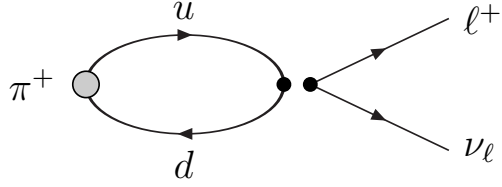}\qquad\qquad
\caption{Correlation function used to calculate the amplitude for the leptonic decay of the pion in pure QCD. The two black filled circles represent the local current-current operator $(\bar{d}\gamma^\mu_Lu)\,(\bar{\nu}_\ell\gamma_\mu\ell)$; the circles are displaced for convenience.\label{fig:LO}}
\end{figure}

In this section we begin our explanation of how the calculations of the amplitudes for the processes $\pi^+\to\ell^+\nu$ and $\pi^+\to\ell^+\nu\gamma$ are to be performed. Before entering into the details however, we discuss more extensively the structure of the different terms appearing in Eq.\,(\ref{eq:master}). 

Since we add and subtract the same perturbative quantity $\Gamma_0^{\mathrm{pt}}$, we find it convenient to choose this to be the virtual decay rate for a point-like pion computed in the W-regularisation. In this way we obtain the important  advantage that the difference of the first two terms ($\Delta\Gamma_0(L)$) and the sum of the last two terms ($\Gamma^\mathrm{pt}(\Delta E)$) on the r.h.s. of Eq.~(\ref{eq:master}) are separately ultraviolet and infrared finite. 

Let $\sqrt{Z_\ell}$ be the contribution to the decay amplitude from the electromagnetic wave-function renormalisation of the final state lepton (see the diagram in Fig.\,\ref{fig:virtualNLO}(d)). An important simplifying feature of this calculation is that $Z_\ell$ cancels in the difference $\Gamma_0-\Gamma_0^\mathrm{pt}$. This is because 
in any scheme and using the same value of the decay constant $f_\pi$, the contribution from the diagram in Fig.\,\ref{fig:virtualNLO}(d) computed non-perturbatively or perturbatively with the point-like approximation are the same. Thus we only need to calculate $Z_\ell$ directly in infinite volume and include it in the second term on the right-hand side of Eq.\,(\ref{eq:master}). As a result of this cancellation it is convenient to rewrite $\Gamma_0$ and $\Gamma_0^\mathrm{pt}$ in the form:
\begin{equation}
\Gamma_0 =  \Gamma_0^{\mathrm{tree}}+ \Gamma_0^{\alpha}+ \Gamma_0^{\mathrm{(d)}}\quad\mathrm{and}\quad
\Gamma_0^{\mathrm{pt}} =  \Gamma_0^{\mathrm{tree}}+ \Gamma_0^{\alpha,\mathrm{pt}}+ \Gamma_0^{\mathrm{(d),pt}}\,,  
\end{equation}
where the superscript {\footnotesize tree} indicates the width in the absence of electromagnetic effects, {\footnotesize $(d)$} denotes the contribution from the leptonic wave function renormalisation and the index {\footnotesize$\alpha$} represents the remaining contributions of $O(\alpha)$ other than those proportional to $Z_\ell$. In this notation the above discussion can be summarised by saying that  $\Gamma_0^{\mathrm{(d)}}=\Gamma_0^{\mathrm{(d),pt}}$ and that the calculation of $\Delta\Gamma_0(L)$ at $O(\alpha)$ reduces to that of computing $\Gamma_0^{\alpha}-\Gamma_0^{\alpha,\mathrm{pt}}$.

Having eliminated the need to include the effects of the lepton's wave-function renormalisation from the evaluation of  $\Delta\Gamma_0(L)$, 
we need to make the corresponding modification in the factor(s) relating the lattice and $W$ regularisations. This simply amounts to subtracting the
term corresponding to the matching between the lattice to $W$ regularisations of the lepton wave function renormalisation diagram.
With the Wilson action (for both gluons and fermions) for example, the $O(\alpha)$ contribution to this matching factor is 
\begin{equation}\label{eq:zlmatching}
\Delta Z_\ell^\mathrm{W-reg} = \frac{\alpha}{4\pi}\left( -\frac{3}{2} - \log a^2M_W^2 - 11.852 \right)\,.  \end{equation}
Thus, with the Wilson action, we can avoid calculating the effects of the lepton's wave-function renormalisation in $\Delta\Gamma_0(L)$ by neglecting the diagram in Fig.\,\ref{fig:virtualNLO}(d) and the corresponding diagram with the point-like pion, and simply replacing $O_1^\mathrm{W-reg}$ in Eq.(\ref{eq:matchingWilson}) by
 \begin{eqnarray} 
\tilde O_1^\mathrm{W-reg}&=&\left(1+\frac{\alpha}{4\pi}\left( \frac52 \, \log a^2M_W^2-8.863 \right)\right)O_1^\mathrm{bare}
+  \frac{\alpha}{4\pi} \,  \left( 0.536\,O_2^\mathrm{bare} \right. \nonumber\\ 
&& \left. \hspace{0.3in}+1.607\,O_3^\mathrm{bare}-3.214\,O_4^\mathrm{bare}-0.804\,O_5^\mathrm{bare} \right) \,.\label{eq:matchingWilsontilde}
\end{eqnarray}
Such matching factors depend, of course, on the lattice discretisation of QCD and we simply present the results for the Wilson action for illustration.

Of course  $\Gamma_0^{\mathrm{(d),pt}}$ needs to be computed for the second term on the right-hand side of Eq.\,(\ref{eq:master}). This is a straightforward perturbative calculation in infinite-volume and gives
\begin{equation}\label{eq:leptonwfpt}
\Gamma_0^{\mathrm{(d),pt}} =   \Gamma_0^{\mathrm{tree}} ~ \frac{\alpha}{4 \pi}\left\{\log\left(\frac{m_\ell^2}{M_W^2}\right)  - 2 \,  \log \left( \frac{m_\gamma^2}{m_\ell^2}\right)-  \frac{9}{2}
\right\}\,,
\end{equation}
where we use the $W$-regularisation for the ultra-violet divergences and have introduced a mass $m_\gamma$ for the photon in order to regulate the infrared divergences. The explicit expression for $\Gamma_0^\mathrm{tree}$ is given in Eq.~(\ref{eq:loG}) below. Using the
$W$-regularisation we naturally work in the Feynman gauge, but note that with $m_\gamma$ as the infrared regulator the result for $Z_\ell$ is generally gauge-dependent. For example, using dimensional regularisation for the ultraviolet divergences and $m_\gamma$ as the infrared regulator leads to a gauge dependent result for this single diagram (gauge invariance is restored of course for $\Gamma^\mathrm{pt}(\Delta E)$).  

In summary therefore, we need to compute the two quantities
\begin{equation} \label{eq:twoquantities}   
\Delta \Gamma_0(L) = \tilde  \Gamma_0^{\alpha} - \Gamma_0^{\alpha,\mathrm{pt}} \qquad\mathrm{and}\qquad \Gamma^\mathrm{pt}(\Delta E) =  \Gamma_0^{\mathrm{tree}}+ \Gamma_0^{\alpha,\mathrm{pt}}+ \Gamma_0^{\mathrm{(d),pt}}+ \Gamma_1(\Delta E)   \, ,\end{equation}
where $\tilde  \Gamma_0^{\alpha}$ corresponds to  $  \Gamma_0^{\alpha}$ using $\tilde O_1^\mathrm{W-reg}$ instead of $ O_1^\mathrm{W-reg}$. 
Note that $ \Delta \Gamma_0(L)$ and $\Gamma^\mathrm{pt}(\Delta E)$ are separately  infrared finite and the result of the calculation of these two quantities  does not depend on the infrared cutoff. In particular, this means that the infrared cutoff  can be chosen in two different ways for the two quantities. We have decided to give a mass to the photon in the perturbative calculation of $\Gamma^\mathrm{pt}(\Delta E) $, whereas for $ \Delta \Gamma_0(L)$  a possible convenient choice is to use the finite volume as the infrared regulator. This will be explained in more detail in Sec.\,\ref{sec:ir}. 

In the following two sections we discuss the calculation of $\Delta \Gamma_0(L)$ and $\Gamma^\mathrm{pt}(\Delta E)$ respectively.

\section{Calculation of $\Delta\Gamma_0(L)$}\label{sec:DGamma0}

In this section we describe the calculation of the first term on the right-hand side of Eq.\,(\ref{eq:master}), $\Delta\Gamma_0(L)$, at $O(\alpha)$. We start however, by briefly recalling the calculation of $\Gamma_0$ at $O(\alpha^0)$, i.e. without electromagnetism.

\subsection{Calculation of $\Gamma_0$ at $O(\alpha^0)$}\label{subsec:pureQCD}

Without electromagnetic corrections we need to compute the correlation function sketched in Fig.\,\ref{fig:LO}, which is a completely standard calculation. Since the leptonic terms are factorized from the hadronic ones, the amplitude is simply given by 
\begin{eqnarray}
\bar{u}_{\nu_\ell\,\alpha}(p_{\nu_\ell})\,(M_0)_{\alpha\beta}\,v_{\ell\,\beta}(p_\ell)&=&\frac{G_F}{\sqrt{2}}V_{ud}^\ast~\langle\,0\,|\,\bar{d}\gamma^\nu\gamma^5\,u\,|\pi^+(p_\pi)\rangle~\big[\bar{u}_{\nu_\ell}(p_{\nu_\ell})\gamma_\nu(1-\gamma^5)\,v_\ell(p_\ell)\big]\nonumber\\ 
&=&\frac{iG_Ff_\pi}{\sqrt{2}}V_{ud}^\ast\,p_\pi^\nu~\big[\bar{u}_{\nu_\ell}(p_{\nu_\ell})\gamma_\nu(1-\gamma^5)\,v_\ell(p_\ell)\big]\label{eq:A0}\,.
\end{eqnarray}
Here $u,d$ in the matrix element represent the quark fields with the corresponding flavour quantum numbers and $u_{\nu_\ell}$ and $v_\ell$ the spinors of the leptons defined by the subscript. The hadronic matrix element, and hence the decay constant $f_\pi$, are obtained in the standard way by computing the correlation function 
\begin{equation}
C_0(t)\equiv\sum_{\vec{x}}~\langle 0\,\,|\Big(\bar{d}(\vec{0},0)\gamma^4 \gamma^5\,u(\vec{0},0)\Big)\,\phi^\dagger(\vec{x},-t)\,|0\rangle\simeq\frac{Z^\phi_0}{2m_\pi^0}e^{-m_\pi^0 t}\,{\cal A}_0\,,\label{eq:C0}
\end{equation}
where $\phi^\dagger$ is an interpolating operator which can create the pion out of the vacuum, $Z^\phi_0\equiv\langle \pi^+(\vec{0}\hspace{1pt})|\phi^\dagger(0,\vec{0}\hspace{1pt})\,|\,0\rangle$ and ${\cal A}_0 \equiv
\langle\,0\,|\,\bar{d}\gamma^4\gamma^5\,u\,|\pi^+(\vec{0}\,)\rangle_0$.  We have chosen to place the weak current at the origin and to create the pion at negative time $-t$, where $t$ and $T-t$ are sufficiently large to suppress the contributions from heavier states and from the backward propagating pions (this latter condition may be convenient but is not necessary). The subscript or superscript $0$ here denotes the fact that the calculation is performed at $O(\alpha^0)$, i.e. in the absence of electromagnetism. $Z^\phi_0$ is obtained from the two-point correlation function of two $\phi$ operators: 
\begin{equation}
C_0^{\phi\phi}(t)\equiv\sum_{\vec{x}}~\langle\,0\,|T\{\phi(\vec{0},0)\,\phi^\dagger(\vec{x},-t)\}\,|\,0\,\rangle
\simeq\frac{(Z_0^{\phi})^2}{2m_\pi^0}\,e^{-m_\pi^0 t}\,.
\end{equation}
For convenience we take $\phi$ to be a local operator (e.g. at $(\vec{x},-t)\,$ in Eq.\,(\ref{eq:C0})), but this is not necessary for our discussion. Any interpolating operator for the pion on the chosen time slice would do equally well.

Having determined ${\cal A}_0$ and hence the amplitude $\bar{u}_{\nu_\ell\,\alpha}(p_{\nu_\ell})(M_0)_{\alpha\beta}\,v_{\ell\,\beta}(p_\ell)$, the $O(\alpha^0)$ contribution to the decay width is readily obtained
\begin{equation}
\Gamma_0^{\mathrm{tree}}(\pi^+\to\ell^+\nu_{\ell})=\frac{G_F^2\,|V_{ud}|^2f_\pi^2}{8\pi}\,m_\pi\,m_\ell^2\left(1-\frac{m_\ell^2}{m_\pi^2}\right)^{\!\!2}\,.  \label{eq:loG}
\end{equation}
In this equation we use the label {\footnotesize tree} to denote the absence of electromagnetic effects since the subscript {\footnotesize $0$} here indicates that there are no photons in the final state.

\subsection{Calculation at $O(\alpha)$}\label{subsec:Oalpha}

\begin{figure}[t]
\includegraphics[width=0.29\hsize]{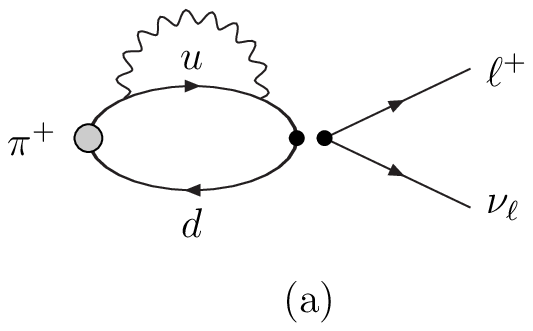}\qquad
\includegraphics[width=0.29\hsize]{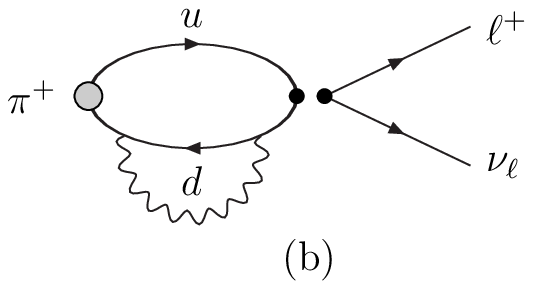}\qquad 
\includegraphics[width=0.29\hsize]{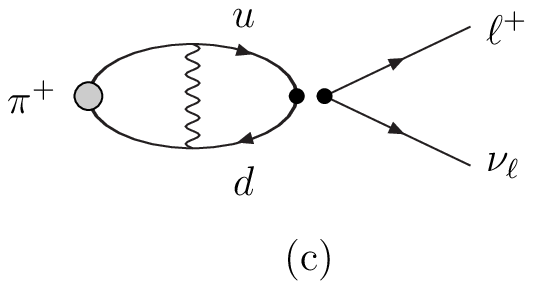}\\[0.3cm]
\includegraphics[width=0.29\hsize]{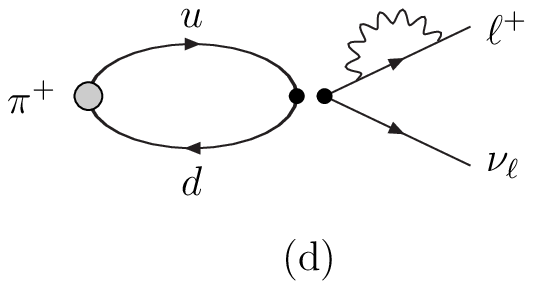}\qquad
\includegraphics[width=0.29\hsize]{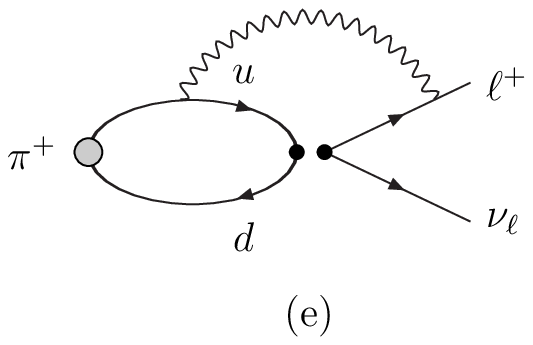}\qquad
\includegraphics[width=0.29\hsize]{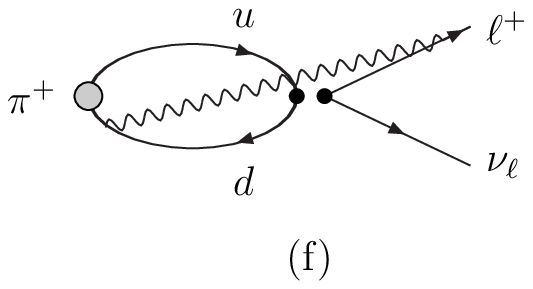}
\caption{Connected diagrams contributing at $O(\alpha)$ contribution to the amplitude for the decay $\pi^+\to\ell^+\nu_l$.\label{fig:virtualNLO}}
\end{figure}

\begin{figure}[t]
\includegraphics[width=0.29\hsize]{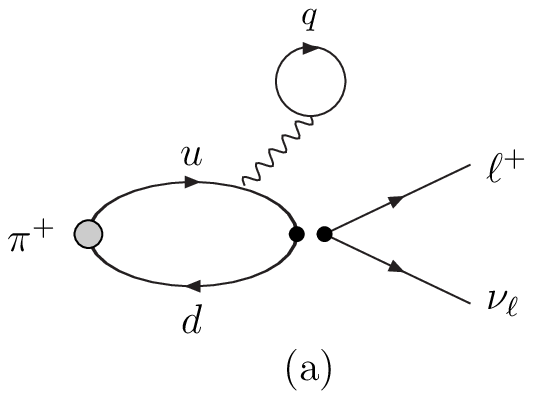}\qquad
\includegraphics[width=0.29\hsize]{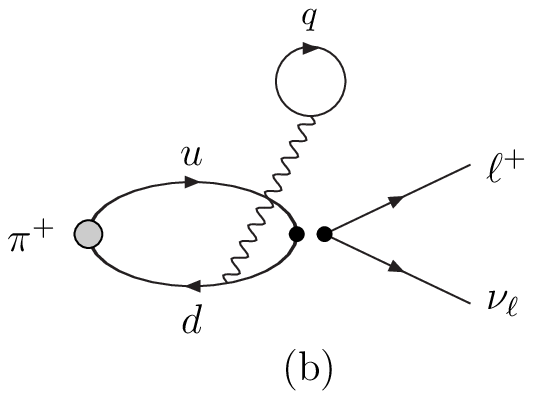}
\qquad \includegraphics[width=0.29\hsize]{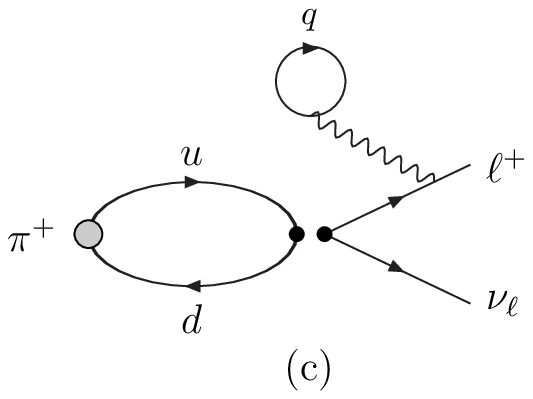}\\[0.3cm]
\includegraphics[width=0.29\hsize]{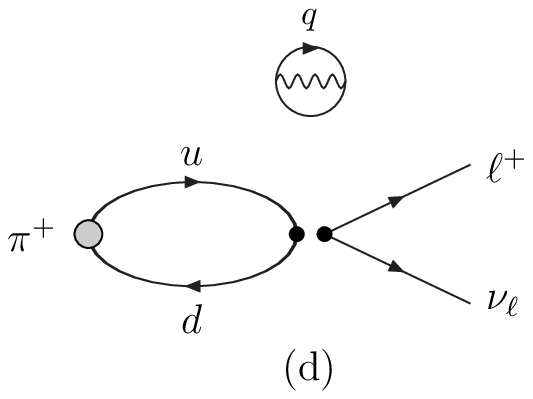}\qquad
\includegraphics[width=0.29\hsize]{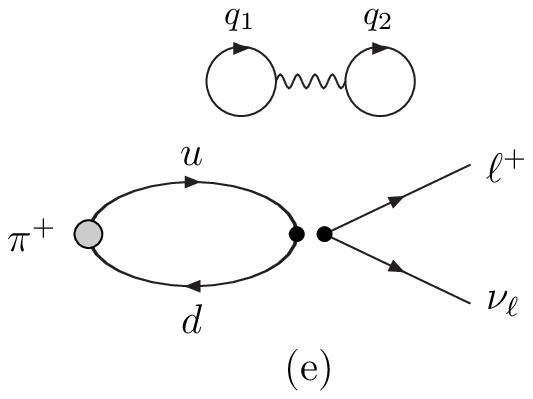}
\caption{Disconnected diagrams contributing at $O(\alpha)$ contribution to the amplitude for the decay $\pi^+\to\ell^+\nu_l$. The curly line represents the photon and a sum over quark flavours $q$, $q_1$ and $q_2$ is to be performed.\label{fig:virtualNLOdisc}}
\end{figure}

We now consider the one-photon exchange contributions to the decay
$\pi^+\to\ell^+\nu_\ell$ and show the corresponding six connected diagrams in Fig.\,\ref{fig:virtualNLO} and the disconnected diagrams in Fig.\,\ref{fig:virtualNLOdisc}. By ``disconnected" here we mean that there is a sea-quark loop connected, as usual, to the remainder of the diagram by a photon and/or gluons (the presence of the gluons is implicit in the diagrams).
The photon propagator in these diagrams in the Feynman gauge and in infinite (Euclidean) volume is given by
\begin{equation}
\delta_{\mu\nu}\Delta(x_1,x_2)=\delta_{\mu\nu}\int\frac{d^4k}{(2\pi)^4}~\frac{e^{ik\cdot(x_1-x_2)}}{k^2}\,.
\end{equation}

In a finite volume the momentum integration is replaced by a summation over the momenta which are allowed by the boundary conditions.
For periodic boundary conditions, we can neglect the contributions from the zero-mode $k=0$ since a very soft photon does not resolve the structure of the pion and its effects cancel in $\Gamma_0-\Gamma_0^\mathrm{pt}$ in Eq.\,(\ref{eq:master}). 
Although we evaluate $\Gamma_0+\Gamma_1(\Delta E)$ (see Eq.~(\ref{eq:Gamma01})) in perturbation theory directly in infinite volume, we note that the same cancellation would happen 
if one were to compute $\Gamma_1(\Delta E)$ also in a finite volume. 
Moreover from a spectral analysis we conclude that such a cancellation also occurs in the Euclidean correlators from which the different contributions to the decay rates are extracted. For this reason in the following $\Gamma_0$ and $\Gamma_0^\mathrm{pt}$ are evaluated separately but using the following expression for the photon propagator in finite volume:
\begin{equation}
\delta_{\mu\nu}\Delta(x_1,x_2)=\delta_{\mu\nu}\,\frac{1}{L^4}\sum_{k=\frac{2\pi}{L}n;\,k\,\neq 0}~\frac{e^{ik\cdot(x_1-x_2)}}{4\sum_\rho\sin^2\frac{k_\rho}{2}}\,,  \label{eq:pp}
\end{equation}
where all quantities are in lattice units and the expression corresponds to the simplest lattice discretisation. $k$, $n$, $x_1$ and $x_2$ are four component vectors and for illustration we have taken the temporal and spatial extents of the lattice to be the same ($L$).

For other quantities, the presence of zero momentum excitations of the photon field is a subtle issue that has to be handled with some care. In the case of the hadron spectrum the problem has been studied in~\cite{Hayakawa:2008an} and, more recently in~\cite{Borsanyi:2014jba,Basak:2014vca}, where 
it has been shown, at $O(\alpha)$,  that the quenching of zero momentum modes corresponds in the infinite-volume limit to the removal of sets of measure zero from the functional integral and that finite volume effects are different for the different prescriptions.

We now divide the discussion of the diagrams in Fig.\,\ref{fig:virtualNLO} and Fig.\,\ref{fig:virtualNLOdisc} into three classes: those in which the photon is attached at both ends to the quarks (diagrams \ref{fig:virtualNLO}(a)-\ref{fig:virtualNLO}(c) and \ref{fig:virtualNLOdisc}(a), (b) and (d)), those in which the photon propagates between one of the quarks and the outgoing lepton (diagrams \ref{fig:virtualNLO}(e), \ref{fig:virtualNLO}(f) and \ref{fig:virtualNLOdisc}(c)) and finally diagram \ref{fig:virtualNLO}(d) which corresponds to the mass and wave-function normalisation of the charged lepton. We have already discussed the treatment of the wave function renormalisation of the lepton in detail in Sec.\,\ref{sec:structure} so we now turn to the remaining diagrams.

\subsubsection{The evaluation of diagrams Fig.\,\ref{fig:virtualNLO}(a)-(c) and    Fig.\,\ref{fig:virtualNLOdisc}(a),(b) and (d)}\label{sec:diagsac}

We start by considering the connected diagrams \ref{fig:virtualNLO}(a)-(c). For these diagrams, the leptonic contribution to the amplitude is contained in the factor $ \big[\bar{u}_{\nu_\ell}(p_{\nu_\ell})\gamma^\nu(1-\gamma^5)\,v_\ell(p_\ell)\big]$ and we need to compute the Euclidean hadronic correlation function
\begin{equation}\label{eq:c1}
C_1(t)=-\frac12\int\,d^3\vec{x}\,d^{\,4}\hspace{-1pt}x_1\,d^{\,4}\hspace{-1pt}x_2~\langle 0|T\big\{J_W^\nu(0)\,j_\mu(x_1)j_\mu(x_2)\phi^\dagger(\vec{x},-t)\big\}\,|\,0\rangle~\Delta(x_1,x_2)\,.
\end{equation}
where $T$ represents time-ordering, $J_W^\nu$ is the $V\textrm{--}A$ current $\bar{d}\gamma^\nu(1-\gamma^5)\,u$ and we take $-t<0$. $j_\mu$ is the hadronic component of the electromagnetic current and we find it convenient to include the charges of the quarks $Q_f$ in the definition of $j$:
\begin{equation}
j_\mu(x)=\sum_f\,Q_f\,\bar{f}(x)\gamma_\mu f(x)\,,
\end{equation}
where the sum is over all quark flavours $f$. 
The factor of $1/2$ is the standard combinatorial one. 

The computations are performed in Euclidean space and in a finite-volume with the photon propagator $\Delta$  given in Eq.\,(\ref{eq:pp}) (or the corresponding expression for other lattice discretisations). 
The absence of the zero mode in the photon propagator implies a gap between $m_\pi$ and the energies of the other eigenstates. Provided one can separate the contributions of these heavier states from that of the pion, one can perform the continuation of the correlation function in Eq.\,(\ref{eq:c1}) from Minkowski to Euclidean space without encountering any singularities. From the correlation function $C_1(t)$ we obtain the electromagnetic shift in the mass of the pion and also a contribution to the physical decay amplitude, as we now explain. For sufficiently large $t$ the correlation function is dominated by the ground state, i.e. the pion, and we have
\begin{equation}\label{eq:c01}
C_0(t)+C_1(t)\simeq \frac{e^{-m_\pi t}}{2m_\pi}\,Z^\phi\,\langle \,0\,|J^0_W(0)\,|\,\pi^+\rangle\,,
\end{equation}
where the electromagnetic terms are included in all factors (up to $O(\alpha)$). Writing $m_\pi=m_\pi^0+\delta m_\pi$, where $\delta m_\pi$ is the $O(\alpha)$ mass shift, 
\begin{equation}
e^{-m_\pi t}\simeq e^{-m_\pi^0 t}\,(1-\delta m_\pi\,t)
\end{equation}
so that $C_1(t)$ is of the schematic form
\begin{equation}
C_1(t)=C_0(t)\,(c_1\,t+c_2)\,.
\end{equation}
By determining $c_1$ we obtain the electromagnetic mass shift, $\delta m_\pi=-c_1$, and from $c_2$ we obtain the electromagnetic correction to $Z^\phi\,\langle \,0\,|J_W(0)\,|\,\pi^+\rangle/2m_\pi$\,. Note that  $\delta m_\pi$ is gauge invariant and infrared finite, whereas the coefficient $c_2$ obtained from these diagrams is neither.

In order to obtain the contribution to the $\pi\to\ell\nu_\ell$ decay amplitude ${\cal A}$ we need to remove the factor $(e^{-m_\pi t}/2m_\pi) Z^\phi$ on the right-hand side of Eq.\,(\ref{eq:c01}), including the $O(\alpha)$ corrections to this factor. Having determined $c_1$, we are in a position to subtract the corrections present in $m_\pi$. The $O(\alpha)$ corrections to $Z^\phi$ are determined in the standard way, by performing the corresponding calculation to $C_1(t)$ but with the axial current $A$ replaced by $\phi$:
\begin{eqnarray}
C_1^{\phi\phi}(t)&=&-\frac12\int\,d^3\vec{x}\,d^4x_1\,d^4x_2~\langle 0|T\big\{\phi(\vec{0},0)\,j_\mu(x_1)j_\mu(x_2)\phi^\dagger(\vec{x},t)\big\}\,|\,0\rangle\,\Delta(x_1,x_2)\\ 
&=&C_0^{\phi\phi}(t)(c_1t+c_2^{\phi\phi})\,.
\end{eqnarray}
We finally obtain
\begin{equation}
Z^\phi=Z^\phi_0\left(1+\frac12\bigg(c_2^{\phi\phi}-\frac{c_1}{m_\pi^0}\bigg)\right)\,,
\end{equation}
and the $O(\alpha)$ contribution to the amplitude from these three diagrams is 
\begin{equation}
\delta{\cal A}={\cal A}_0\,\bigg(c_2-\frac{c_2^{\phi\phi}}{2}-\frac{c_1}{2m_\pi^0}\bigg)\,.
\end{equation}
For these three diagrams the $O(\alpha)$ term can be simply considered as a correction to $f_\pi$. Note however, that such an ``$f_\pi$" would not be a physical quantity as it contains infrared divergences.

The treatment of the disconnected diagrams in Figs.\,\ref{fig:virtualNLOdisc}(a), (b) and (d) follows in exactly the same way. These diagrams contribute to the electromagnetic  corrections to both the pion mass and the decay amplitude in an analogous way to the discussion of the connected diagrams above . It is standard and straightforward to write down the corresponding correlation functions in terms of quark propagators. We do not discuss here the different possibilities for generating the necessary quark propagators to evaluate the diagrams; for example we can imagine using sequential propagators or some techniques to generate all-to-all quark propagators.

\subsubsection{The evaluation of diagrams Fig.\,\ref{fig:virtualNLO}(e)-(f)}

For these diagrams the leptonic and hadronic contributions do not factorise and indeed the contribution cannot be written simply in terms of the parameter $f_\pi$. We start by considering the Minkowski space quantity
\begin{eqnarray}\label{eq:diagsef1}
\bar{u}_{\nu_\ell\,\alpha}(p_{\nu_\ell})(\bar M_1)_{\alpha\beta}\,v_{\ell\,\beta}(p_\ell)&=&
-\int\!d^{\,4}\hspace{-1pt}x_1\,d^{\,4}\hspace{-1pt}x_2\, \bra{0}\,T(j_\mu(x_1) J^\nu_W(0))\,\ket{\pi}
 \\
&&\hspace{-1in}\times
\,i D_M(x_1,x_2)
\big\{\bar{u}_{\nu_\ell}(p_{\nu_\ell}) \gamma^\nu(1-\gamma^5)(i S_M(x_2))\gamma^\mu v_\ell(p_\ell)\big\}e^{ip_\ell\cdot x_2}\,, \nonumber
\end{eqnarray}
where $iS_M$ and $iD_M$ are the lepton and (Feynman gauge) photon propagators respectively in Minkowski space (more precisely the photon propagator with Lorentz indices 
$(\rho,\sigma)$ is $iD_Mg_{\rho\sigma}$, but  the Lorentz indices have been contracted with the electromagnetic currents in (\ref{eq:diagsef1})). In order to demonstrate that we can obtain the $O(\alpha)$ corrections to the decay amplitude from a Euclidean space correlation function,  
we use the reduction formula to rewrite the expression in Eq.\,(\ref{eq:diagsef1}) as
\begin{eqnarray}  
&&\bar{u}_{\nu_\ell\,\alpha}(p_{\nu_\ell})(\bar M_1)_{\alpha\beta}\,v_{\ell\,\beta}(p_\ell)=
i \lim_{k_0 \rightarrow m_\pi} ({k_0}^2 - m_\pi^2) 
\int d^4x_1\, d^4x_2\, d^4x\, e^{-ik^0 x^0}\,  
\nonumber\\ 
&&\hspace{-0.3in}
\bra{0}T(j_\mu(x_1) J^\nu_W (0) \pi (x)) \ket{0}\,
i D_M(x_1,x_2)
  \big[\bar{u}_{\nu_\ell}(p_{\nu_\ell}) \gamma_{\nu}(1-\gamma^5)(i S_M(x_2))\gamma^\mu v_\ell(p_\ell)\big]
  e^{ip_\ell\cdot x_2}\, ,\label{eq:reduction}
\end{eqnarray}
where $\pi(x)$ is the field which creates a pion with amplitude 1. On the other hand
the Euclidean space correlation function which we propose to compute is 
\begin{eqnarray}
\bar C_1(t)_{\alpha\beta}&=&-\int\! d^3\vec{x}\,d^4x_1\,d^4x_2~\langle 0|T\big\{J^\nu_W(0)\,j_\mu(x_1)\phi^\dagger(\vec{x},-t)\big\}|\,0\rangle~\Delta(x_1,x_2)\nonumber\\ 
&&\hspace{0.7in}\times \big(\gamma_\nu(1-\gamma^5)S(0,x_2)\gamma_\mu\big)_{\alpha\beta}\, e^{E_\ell\,t_2}e^{-i\vec{p}_\ell\cdot \vec{x}_2}.\label{eq:c2}
\end{eqnarray}
Here $S$ and $\Delta$ are Euclidean propagators, and $\alpha,\beta$ are spinor indices.  
Similarly to the discussion in Sec.~\ref{sec:diagsac}, provided that the pion is the lightest hadronic state then for large $t$, $\bar C_1(t)$ is dominated by the matrix element with a single pion in the initial state. 

\begin{figure}[t]
\includegraphics[width=0.25\hsize]{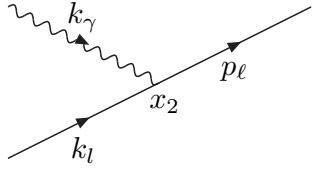}\qquad\qquad
\caption{Zoom of the lepton-photon vertex at $x_2$ from the diagrams in Fig.\,\ref{fig:virtualNLO}(e) and (f). \label{fig:llgamma}}
\end{figure}

In view of the factor $e^{E_\ell\,t_2}$ on the right-hand side of Eq.\,(\ref{eq:c2}), the
new feature in the evaluation of the diagrams in Fig.\,\ref{fig:virtualNLO}\,(e) and (f) is that we need to ensure that the $t_2$ integration converges as $|t_2|\to\infty$. For $t_2<0$ the convergence of the integral is improved by the presence of the exponential factor and so we limit the discussion to the case $t_2\to\infty$. $E_\ell=\sqrt{m_\ell^2+\vec{p}_\ell^{~2}}$ is the energy of the outgoing charged lepton with three-momentum $\vec{p}_\ell$. To determine the $t_2\to\infty$ behaviour, consider the lepton-photon vertex at $x_2$ from the diagrams in Fig.\,\ref{fig:virtualNLO}(e) and (f), redrawn in Fig.\,\ref{fig:llgamma}. $k_\ell$ and $k_\gamma$ are the four-momentum variables in the Fourier transform of the propagators $S(x_2)$ and $\Delta(x_1,x_2)$ respectively in Eqs.\,(\ref{eq:diagsef1})\,-\,(\ref{eq:c2}).
The $t_2$ integration is indeed convergent as we now show explicitly.\\
1. The integration over $\vec{x}_2$ implies three-momentum conservation at this vertex so that in the sum over the momenta $\vec{k}_\ell+\vec{k}_\gamma=\vec{p}_l$, where $p_\ell$ is the momentum of the outgoing charged lepton.\\ 
2. The integrations over the energies $k_{4\,\ell}$ and $k_{4\,\gamma}$ lead to the exponential factor $e^{-(\omega_\ell+\omega_\gamma)t_2}$, where $\omega_\ell=\sqrt{\vec{k}_\ell^{\,2}+m_\ell^2}$, $\omega_\gamma=\sqrt{\vec{k}_\gamma^{\,2}+m_\gamma^2}$, and $m_\gamma$ is the mass of the photon introduced as an infra-red cut-off. The large $t_2$ behaviour is therefore given by the factor 
$e^{-(\omega_\ell+\omega_\gamma-E_\ell)t_2}$.\\ 
3. A simple kinematical exercise shows that in the sum over $\vec{k}_\gamma$ (with $\vec{k}_\ell=\vec{p}_\ell-\vec{k}_\gamma$), the minimum value of $\omega_\ell+\omega_\gamma$ is given by
\begin{equation}
(\omega_\ell+\omega_\gamma)_\mathrm{min}=\sqrt{(m_\ell+m_\gamma)^2+\vec{p}_\ell^{~2}}\,.
\end{equation}
4. Thus for non-zero $m_\gamma$, the exponent in $e^{-(\omega_\ell+\omega_\gamma -E_\ell)t_2}$ for large $t_2$ is negative for every term in the summation over $k_\gamma$ and the integral over $t_2$ is convergent so that the continuation from Minkowski to Euclidean space can be performed.\\ 
5. We note that the integration over $t_2$ is also convergent if we set $m_\gamma=0$ but remove the $\vec{k}=0$ mode in finite volume. In this case $\omega_\ell+\omega_\gamma>E_l+[1-(p_\ell/E_\ell)]|\vec{k}_\mathrm{min}|$.

In summary the $t_2$ integration is convergent because for every term in the sum over momenta $\omega_\ell+\omega_\gamma>E_l$ and so for sufficiently large $t$ we can write 
\begin{equation}\label{eq:c2M}
\bar C_1(t)_{\alpha\beta}\simeq Z^\phi_0\,\frac{e^{-m_\pi^0t}}{2m_\pi^0}\,(\bar M_1)_{\alpha\beta}
\end{equation}
and the contribution from the diagrams of Fig.\,\ref{fig:virtualNLO}(e) and \ref{fig:virtualNLO}(f) is $\bar{u}_\alpha(p_{\nu_\ell})(\bar M_1)_{\alpha\beta}v_\beta(p_\ell)$.
This completes the demonstration that the Minkowski-space amplitude (\ref{eq:reduction}) is equal to the pion contribution to the Euclidean correlation function (\ref{eq:c2}), up to a factor $Z_0^\phi$ which accounts for the normalisation of the pion field.

Again the evaluation of the correction to the amplitude from the disconnected diagram in Fig.\,\ref{fig:virtualNLOdisc}(c) follows in an analogous way.

\section{Calculation of $\Gamma^\mathrm{pt}(\Delta E)$}\label{sec:Gamma1}

The evaluation in perturbation theory of  the total width $\Gamma^\mathrm{pt}= \Gamma_0^{\mathrm{pt}}+\Gamma_1^\mathrm{pt}$ in infinite volume, was performed by Berman  and  Kinoshita in 1958/9~\cite{Berman:1958ti,Kinoshita:1959ha}, using the Pauli-Villars regulator for the ultraviolet divergences and a photon mass to regulate the infrared divergences in both $\Gamma_0^{\mathrm{pt}}$ and $\Gamma_1^\mathrm{pt}$.  $\Gamma_1^\mathrm{pt}$ is the rate for process $\pi^+\to\ell^+\nu_\ell\,\gamma$ for a pointlike pion with the energy of the photon integrated over the full kinematic range. We have added the label {\footnotesize pt} in $\Gamma_1^\mathrm{pt}$ to remind us that
the integration includes contributions from regions of phase space in which the photon is not sufficiently soft for the structure of the pion to be reliably neglected. We do not include this label when writing $\Gamma_1(\Delta E)$ because we envisage that $\Delta E$ is sufficiently small so that the pointlike approximation reproduces the full calculation.

In our calculation,  $\Gamma_0^{\mathrm{pt}}$ is evaluated in the W-regularisation, so that the ultra-violet divergences are replaced by logarithms of $M_W$.
For convenience we rewrite here the expression for $ \Gamma^\mathrm{pt}(\Delta E)$ from Eq.\,(\ref{eq:twoquantities})
\begin{eqnarray} \Gamma^\mathrm{pt}(\Delta E) =  \Gamma_0^{\mathrm{pt}} + \Gamma_1(\Delta E) = \Gamma_0^{\mathrm{tree}}+ \Gamma_0^{\alpha,\mathrm{pt}}+ \Gamma_0^{\mathrm{(d),pt}}+ \Gamma_1(\Delta E)\,. \label{eq:gammadeltaf} \end{eqnarray} 
 $\Gamma_0^{\mathrm{tree}}$ and $ \Gamma_0^{\mathrm{(d)},\mathrm{pt}}$  have already been  presented  in Eqs.~(\ref{eq:loG}) and (\ref{eq:leptonwfpt}) respectively.  In the  following we give separately  the results of the remaining contributions to $\Gamma^\mathrm{pt}(\Delta E)$ also using a photon mass $m_\gamma$  as the  infrared regulator. We  neglect powers of $m_\gamma$ in all the results. 

In the perturbative calculation we use the following Lagrangian for the interaction of a point-like pion with the leptons:
\begin{eqnarray}
\mathcal{L}_{\pi-\ell-\nu_\ell} &=&i\, G_F f_\pi  V_{ud}^* ~  \left\{(\partial_\mu -ieA_\mu) \pi\right\}  \, \left\{\bar{\psi}_ {\nu_\ell} \frac{1+\gamma_5}{2} \gamma^\mu \psi_\ell\right\}  + 
\mathrm{Hermitian~conjugate}\,.   
\label{eq:covdervertex}\end{eqnarray}
The  corresponding Feynman rules  are:
\begin{eqnarray}
\includegraphics[width=0.5\hsize]{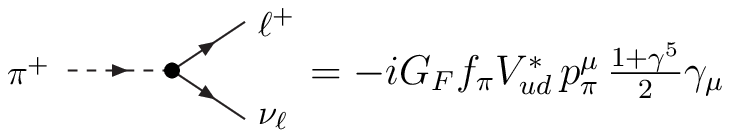}\label{eq:covdervertexrule}\\
\includegraphics[width=0.5\hsize]{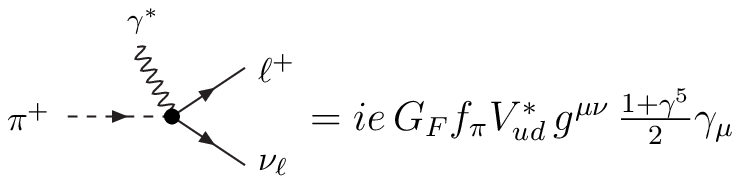}\nonumber
\end{eqnarray}
In addition we have used the standard Feynman rules of scalar electromagnetism for the interactions of charged pions in an electromagnetic field.

\begin{figure}
\includegraphics[width=0.6\hsize]{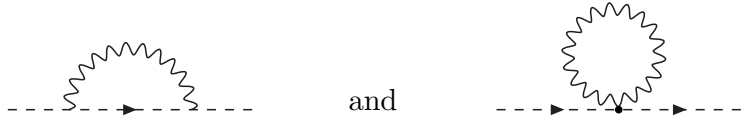}
\caption{One loop diagrams contributing to the wave-function renormalisation of a point-like pion.\label{fig:freelself}} 
\end{figure} 
We start by giving the $O(\alpha)$ contributions to $\Gamma_0^{\alpha,\mathrm{pt}}$.\\ 
\noindent $\bullet$ {\it Wave function renormalisation of the pion}: The contribution of the pion wave function renormalisation to $\Gamma_0^{\alpha,\mathrm{pt}}$ is obtained from the diagrams in Fig.\,\ref{fig:freelself}
and is given by
\begin{equation}  
\Gamma_0^{\mathrm{\pi} }= \Gamma_0^{\mathrm{tree}} \times \frac{\alpha}{4 \pi} \,  Z_\pi\,,\quad\mathrm{where}\quad   
 Z_\pi = - 2 \log\left(\frac{m_\pi^2}{M_W^2}\right)  - 2 \log\left(\frac{m_\gamma^2}{m_\pi^2}\right)  -\frac{3}{2}  \, .
\label{eq:zpi}  \end{equation}
These diagrams correspond to those in  Fig.~\ref{fig:virtualNLO}(a), Fig.~\ref{fig:virtualNLO}(b) and Fig.~\ref{fig:virtualNLO}(c) in the composite case.

\begin{figure}[t]
\includegraphics[width=0.295\hsize]{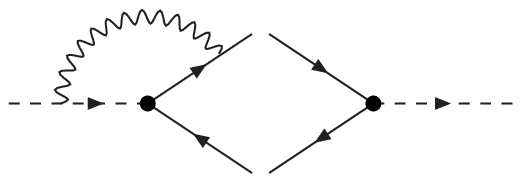}\qquad\includegraphics[width=0.295\hsize]{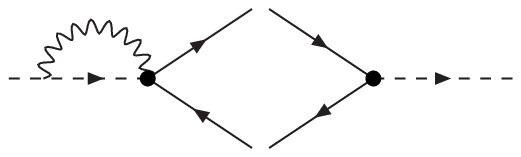}
\qquad\includegraphics[width=0.295\hsize]{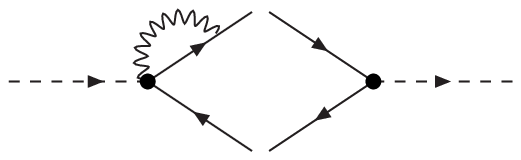}
\caption{Radiative corrections to the pion-lepton vertex. The diagrams represent $O(\alpha)$ contributions to $\Gamma_0^\mathrm{pt}$. The left part of each diagram represents a contribution to the amplitude and the right part the tree-level contribution to the hermitian conjugate of the amplitude. The corresponding diagrams containing the radiative correction on the right-hand side of each diagram are also included.\label{fig:kinod}}
\end{figure}
\noindent $\bullet$~{\it $\pi$\,-\,$\ell$ Vertex}:  The remaining graphs contributing to $\Gamma_0^{\alpha,\mathrm{pt}}$ are the $\pi$\,-\,$\ell$ vertex corrections from the diagrams shown in Fig.\,\ref{fig:kinod} and their complex conjugates.
The contribution from these diagrams is
\begin{eqnarray}  \Gamma_0^{\mathrm{\pi-\ell} }&=&   \Gamma_0^{\mathrm{tree}} \times \frac{\alpha}{4 \pi} \,  Z_{\pi-\ell}\quad\mathrm{where}\\     \label{eq:pionl}  
 &&\hspace{-1in}Z_{\pi-\ell}  = -2 \frac{1+ r_\ell^2}{1- r_\ell^2} \,  \log\left( r_\ell^2\right) \,  \log\left(\frac{m_\gamma^2}{m_\pi^2}\right)  + 4 \log\left(\frac{m_\pi^2}{M_W^2}\right) 
+ \nonumber\\ &&\hspace{0.5in} \frac{1+ r_\ell^2}{1- r_\ell^2} \,  \log^2\left( r_\ell^2\right)  +  2 \frac{1-3 r_\ell^2}{1- r_\ell^2} \,  \log\left( r_\ell^2\right)  -1    \, ,  \label{eq:zpil}  \end{eqnarray}
and $r_\ell = m_\ell/m_\pi$.
These diagrams correspond to the diagrams  Fig.~\ref{fig:virtualNLO}(e)  and  Fig.~\ref{fig:virtualNLO}(f)  in the composite pion case.

Next we give the contributions to $\Gamma_1(\Delta E)$ where the real photon is emitted and absorbed by the pion ($\pi\pi$), the charged lepton ($\ell\ell$) or emitted by the pion and absorbed by the lepton or vice-versa ($\pi\ell$). The results are presented in the Feynman gauge:
\begin{equation}
\sum_r \varepsilon^\star_\mu(k,r)\, \varepsilon_\nu(k,r) = g_{\mu\nu} \;,
\end{equation}
where $\varepsilon_\mu(k,r)$ are the polarisation vectors of the real photon carrying a momentum $k$, with $k^2=0$ in Minkowski space.

\begin{figure}
\includegraphics[width=0.29\hsize]{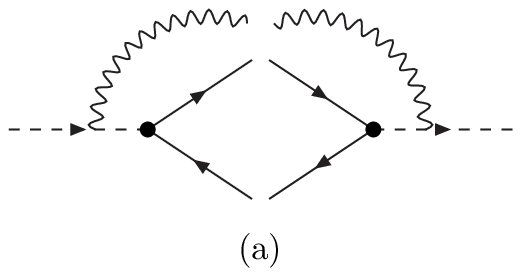}\qquad\includegraphics[width=0.29\hsize]{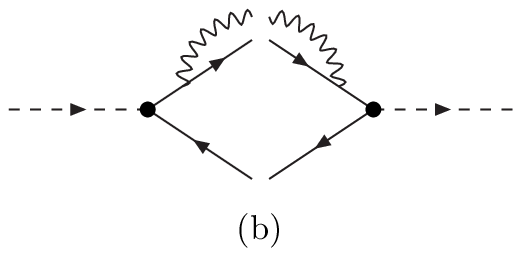}
\qquad\includegraphics[width=0.29\hsize]{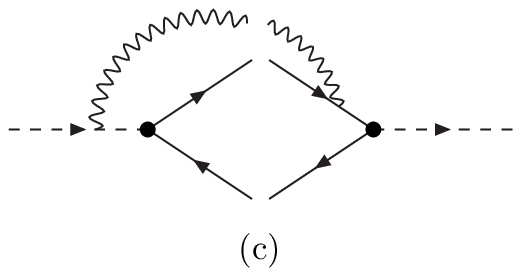}\\[0.2in]
\includegraphics[width=0.29\hsize]{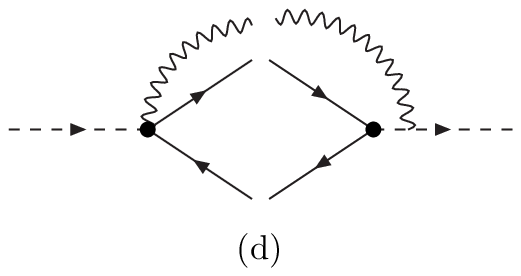}\qquad\includegraphics[width=0.29\hsize]{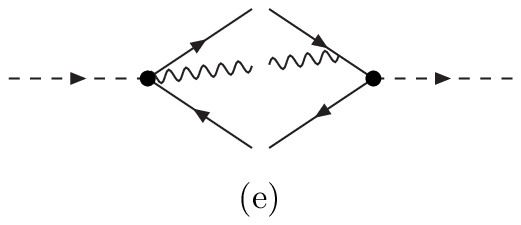}
\qquad\includegraphics[width=0.29\hsize]{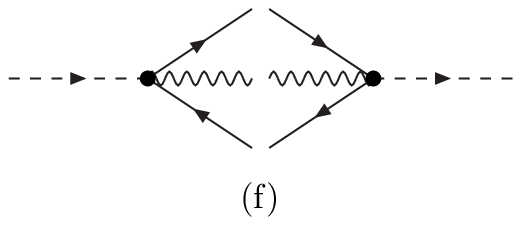}
\caption{Diagrams contributing to $\Gamma_1(\Delta E)$. For diagrams (c), (d) and (e) the ``conjugate" contributions in which the photon vertices on the left and right of each diagram are interchanged are also to be included.\label{fig:realpt}}
\end{figure}

\noindent$\bullet$~{\it Real photon emission,  $\pi \pi$}:  The contribution to $\Gamma_1(\Delta E)$ from the emission and absorption of a real photon from the pion, represented by diagram (a) in Fig.\,\ref{fig:realpt},
is given by 
\begin{eqnarray}  \hspace{1in}\Gamma_1^{\pi\pi} &=&   \Gamma_0^{\mathrm{tree}} \times \frac{\alpha}{4 \pi} \,\left( R_1^{\pi\pi}+ R_2^{\pi\pi} \right) \, ,\quad\mathrm{where}   \label{eq:pionpion} 
\\ \nonumber\\    
&&\hspace{-1.8in}R_1^{\pi\pi}=  2 \log\left(\frac{m_\gamma^2}{4\Delta E^2}\right) +4\,,\quad R_2^{\pi\pi}= 
\frac{2r_\ell^4}{(1-r_\ell^2)^2}\, \log(1-r_E)
+\frac{r_E\left( 6-r_E-4r_\ell^2\right)}{(1-r_\ell^2)^2}
  \; ,
\end{eqnarray}
 $r_E = 2 \Delta E/m_\pi$ and $0 \le r_E \le 1-r_\ell^2$. Here we have separated $R_1^{\pi\pi}$, the contribution in the eikonal approximation from $R_2^{\pi\pi}$ which vanishes as $\Delta E \to 0$. In the eikonal approximation only the leading terms in the photon's momenta are kept in the numerator and denominator of the integrand as $r_E\to 0$. $R_1^{\pi\pi}$ contains the infrared divergence.
 
\noindent$\bullet$~{\it Real photon emission,  $\ell\ell$}:
The contribution
to $\Gamma_1(\Delta E)$ from the emission and absorption of a real photon from the charged lepton, represented by the diagram (b) in Fig.\,\ref{fig:realpt},
is given by 
\begin{eqnarray}  \Gamma_1^{ \ell \ell} =   \Gamma_0^{\mathrm{tree}} \times \frac{\alpha}{4 \pi} \,\left( R_1^{ \ell \ell}+ R_2^{ \ell \ell} \right) \, ,\quad\mathrm{where}   \label{eq:leplep}  \end{eqnarray}
\begin{eqnarray}  
R_1^{\ell\ell}&=&
2 \log\left(\frac{m_\gamma^2}{4\Delta E^2}\right) 
-2\frac{1+r_\ell^2}{1-r_\ell^2} \log(r_\ell^2)
 \; ,\quad\mathrm{and}
\nonumber \\
\nonumber \\
R_2^{\ell\ell}&=& 
\frac{r_E^2-1 +  (4 r_E-6) r_\ell^2}{(1-r_\ell^2)^2}\ \log(1-r_E)
-\frac{r_E(r_E + 4 r_\ell^2)}{(1-r_\ell^2)^2}\ \log(r_\ell^2)\nonumber\\ 
&&\hspace{1in}+\frac{r_E(6 - 3 r_E - 20 r_\ell^2)}{2(1-r_\ell^2)^2}
\; .\end{eqnarray}
  
\noindent$\bullet$~{\it Real photon emission, $ \pi \ell$}:  Finally, the contribution
to $\Gamma_1(\Delta E)$ from the emission of a real photon from the pion and its absorption by the charged lepton, represented by the diagrams (c)\,--\,(f) in Fig.\,\ref{fig:realpt},
is given by 
\begin{eqnarray}  \Gamma_1^{ \pi \ell} =   \Gamma_0^{\mathrm{tree}} \times \frac{\alpha}{4 \pi} \,\left( R_1^{ \pi \ell}+ R_2^{ \pi \ell} \right) \, ,   \label{eq:rleplep}  \end{eqnarray}
where \begin{eqnarray}  
R_1^{\pi\ell}&=&
2\frac{1+r_\ell^2}{1-r_\ell^2}\log(r_\ell^2)
 \log\left(\frac{m_\gamma^2}{4\Delta E^2}\right) 
-\frac{1+r_\ell^2}{1-r_\ell^2}\left[\log(r_\ell^2)\right]^2
-4\frac{1+r_\ell^2}{1-r_\ell^2}\,\mbox{Li}_2(1-r_\ell^2)
 \quad\mathrm{and}
 \nonumber \\
\nonumber \\
R_2^{\pi\ell}&=& 
-2\, \frac{2 r_E + r_\ell^4-2}{(1-r_\ell^2)^2}\,\log(1-r_E)
+\frac{4r_E}{(1-r_\ell^2)^2}\,\log(r_\ell^2)
+\frac{r_E(2 + r_E)}{(1-r_\ell^2)^2}
-4\,\frac{1+r_\ell^2}{1-r_\ell^2}\ \mbox{Li}_2(r_E)
\,.
\nonumber \\
\label{eq:Rpiell}
 \end{eqnarray}
Note that for diagrams (c), (d) and (e) we include the conjugate contribution in which the photon vertices are interchanged between the left and right parts of the diagrams. Thus for example, in addition to diagram (c) there is the diagram in which the photon is emitted from the lepton on the left and absorbed on the pion on the right.

We are now in a position to combine the results in Eqs.\,(\ref{eq:zpi})\,--\,(\ref{eq:Rpiell}) in order to obtain the 
final expression for $\Gamma^\mathrm{pt}(\Delta E)$. As expected the infrared cutoff cancels and we find
\begin{eqnarray}
\Gamma^\mathrm{pt}(\Delta E) &=& \Gamma_0^{\mathrm{tree}} \times  \left( 1+ 
\frac{\alpha}{4\pi}\ \Bigg\{
 3 \log\left(\frac{m_\pi^2}{M_W^2}\right)+  \log\left(r_\ell^2\right) - 4 \log(r_E^2) +\frac{2- 10 r_\ell^2}{1-r_\ell^2} \log(r_\ell^2)  \right.
\nonumber \\
\nonumber \\
&&\qquad\qquad\qquad-2\frac{1+r_\ell^2}{1-r_\ell^2}\ \log(r_E^2)\log(r_\ell^2) -4\frac{1+r_\ell^2}{1-r_\ell^2}\ \mbox{Li}_2(1-r_\ell^2)- 3
\nonumber \\
\nonumber \\
&&
+\Big[\frac{3 + r_E^2 - 6 r_\ell^2  + 4  r_E (-1 +  r_\ell^2) }{(1-r_\ell^2)^2}\ \log(1-r_E)+\frac{r_E (4 - r_E -4 r_\ell^2)}{(1-r_\ell^2)^2}\ \log(r_\ell^2)
\nonumber \\
\nonumber \\
&& 
\left.
\qquad\qquad
-\frac{r_E (-22 + 3  r_E +28  r_\ell^2)}{2(1-r_\ell^2)^2}
-4\frac{1+r_\ell^2}{1-r_\ell^2}\ \mbox{Li}_2(r_E)\Big]\
\Bigg\}  \right) \; .
\label{eq:formulafinal}
\end{eqnarray}
Note that the terms in square brackets in eq.~(\ref{eq:formulafinal}) vanish when $r_E$ goes to zero; in this limit  $\Gamma^\mathrm{pt}(\Delta E)$ is given by its eikonal approximation.

The total rate is readily computed by setting $r_E$ to its maximum value, namely $r_E=1-r_\ell^2$, giving 
\begin{eqnarray}
\Gamma^\mathrm{pt} &=& \Gamma_0^{\mathrm{tree}} \times  \Bigg\{  1+ 
\frac{\alpha}{4\pi}\ \left(
 3 \log\left(\frac{m_\pi^2}{M_W^2}\right) - 8 \log(1-r_\ell^2) -\frac{3 r_\ell^4}{(1-r_\ell^2)^2} \log(r_\ell^2)  \right.  \label{eq:formulafinalI}
\\
&&\hspace{-0.35in}\left. - 8 \frac{1+r_\ell^2}{1-r_\ell^2}\ \mbox{Li}_2(1-r_\ell^2)+\frac{13-19 r_\ell^2}{2(1-r_\ell^2)}  +  \frac{6 - 14 r_\ell^2 - 4(1+r_\ell^2)\log(1- r_\ell^2) }{1-r_\ell^2}\ \log(r_\ell^2) 
  \right) \Bigg\} \; . \nonumber 
\end{eqnarray}
The result in Eq.~(\ref{eq:formulafinalI})  agrees with  the well known  results  in literature~\cite{Berman:1958ti,Kinoshita:1958ru}, which provides an important  check of our calculation. We believe that the result  in  Eq.\,(\ref{eq:formulafinal}) is new.

In the description of our method above, we limit the photon's energy to be smaller than $\Delta E$ to ensure that the photon is sufficiently soft for the pointlike approximation to be valid in the evaluation of $\Gamma_1(\Delta E)$. It is of course possible instead to impose a cut-off on the energy of the final-state lepton, requiring it to be close to its maximum value $E^{\mathrm{max}}_\ell = \frac{m_\pi}{2}(1+r_\ell^2)$. For completeness we also 
give, up to $O(\Delta E_\ell)$,  the distribution for $\Gamma^\mathrm{pt}(\Delta E_\ell)$ defined as 
\begin{eqnarray}
\Gamma^\mathrm{pt}(\Delta E_\ell) = \int^{E^{max}_\ell}_{E^{max}_\ell - \Delta E_\ell } \, dE^\prime  \, \frac{d\Gamma^\mathrm{pt}}{d E^\prime_\ell} \, , \end{eqnarray}
where $0\le \Delta E_\ell \le (m_\pi-m_\ell)^2/(2 m_\pi)$;
\begin{eqnarray}&&
\Gamma^\mathrm{pt}(\Delta E_\ell)  = \Gamma_0^{\mathrm{tree}} \times  \Bigg\{  1+ 
\frac{\alpha}{4\pi}\, \bigg[
 3 \log\left(\frac{m_\pi^2}{M_W^2}\right) + 8 \log\left( 1-r_\ell^2\right) - 7 \nonumber  \\ 
&& \hspace{0.3in}+  \log\left( r_\ell^2\right)   
\frac{3 - 7 r_\ell^2 + 8 \Delta E_\ell   + 4 \left(1+r_\ell^2\right) \log\left(1-r_\ell^2\right)}{1-r_\ell^2} \label{eq:formuladel}  \\ 
&& \hspace{1in}+\log\left( 2  \Delta E_\ell \right) \, \left( -8 -4  \frac{1+r_\ell^2}{1-r_\ell^2}\log\left(r_\ell^2\right)    \right) 
\bigg] \Bigg\} \,. \nonumber 
\end{eqnarray}

\section{Regularisation and cancellation of infrared divergences in finite-volumes}\label{sec:ir}

In the previous section we have explicitly demonstrated the cancellation of infrared divergences in the perturbative quantity $\Gamma^\mathrm{pt}(\Delta E)$. This of course is simply the standard Bloch-Nordsieck cancellation~\cite{Bloch:1937pw}.
In this section we discuss in more detail the cancellation of infrared divergences in
 \begin{eqnarray}    \Delta \Gamma_0(L) = \tilde  \Gamma_0^{\alpha} - \Gamma_0^{\alpha,\mathrm{pt}} \, . \label{eq:51} \end{eqnarray} 
We have already explained in Sec.\,\ref{sec:structure} that the contribution of the lepton's wave function renormalisation in $\Delta\Gamma_0(L)$ is simply to introduce the tilde in $\tilde  \Gamma_0^{\alpha}$, denoting that the corresponding contribution to the matching factor between the lattice and $W$-regularisations is to be removed. We also do not discuss further the evaluation of the remaining infrared-finite terms in the matching factor because these are straightforward to evaluate (see e.g. Eq.\,(\ref{eq:matchingWilsontilde}) for the Wilson action).
Here  we concentrate on the remaining diagrams in Figs.~\ref{fig:virtualNLO} and \ref{fig:virtualNLOdisc} and the corresponding diagrams for  the  point-like meson.  

Although the right-hand side of Eq.\,(\ref{eq:51}) is a difference of decay widths, since at this order the widths are linear in the $O(\alpha)$ virtual amplitude, we can equivalently consider the difference of the $O(\alpha)$ contributions to the amplitudes. 
In order to reduce statistical fluctuations when performing the sum over the gauge field configurations, we define  the ratios
\begin{eqnarray} R^{\alpha}  = \frac{\tilde{\cal A}^\alpha}{{\cal A}_0} \, , \quad \quad R^{\alpha,\mathrm{pt}}  = \frac{{\cal A}^{\alpha,\mathrm{pt}}}{{\cal A}_0} \,  , \end{eqnarray} 
where $\tilde{\cal A}^\alpha$ and ${\cal A}^{\alpha,\mathrm{pt}}$  are the $O(\alpha)$ amplitudes corresponding to the widths in Eq.\,(\ref{eq:51}).  The  non-perturbative   amplitude $\tilde{\cal A}^\alpha$ is  precisely the quantity that we propose to compute numerically in a lattice simulation. It is then combined with ${\cal A}^{\alpha,\mathrm{pt}}$, for which we have given the explicit expression in infinite volume in Sec.\,\ref{sec:Gamma1}. 

In the calculation  of ${\cal A}^{\alpha,\mathrm{pt}}$ we set the mass of the photon to zero and consider the theory on a finite volume of length $L$, which will be used as an infrared regulator. The form of the vertices and   propagators is  the same as in the infinite volume (the ultraviolet cutoff is provided by the W-regularisation), but the momenta are quantized $k_\mu   = 2 \pi /L \times  n_\mu = 2 \pi / (Na)  \times  n_\mu $ where $-\infty \le n_\mu  \le +\infty$ and $N$ is the number of lattice sites in one direction, which for simplicity we take to be the same in all directions. 

The calculation of $\tilde  {\cal A}_0^{\alpha}$  is performed non perturbatively  on the same  finite  volume  as  in the perturbative case, but in a numerical simulation and with  the photon propagator  defined as in Eq.\,(\ref{eq:pp}), which does not contain the zero mode. Indeed, as already discussed in Sec.~\ref{subsec:Oalpha}, the zero mode does not contribute to the difference
\begin{eqnarray} \Delta R(L) =R^{\alpha}-R^{\alpha,\mathrm{pt}}\,. \end{eqnarray}
This is a gauge invariant, ultraviolet and infrared finite quantity and for these reasons we expect that its finite volume effects are comparable to those affecting the $O(\alpha)$ corrections to the hadron masses (that are also gauge invariant, ultraviolet and infrared finite). 
The formalism introduced in this paper was necessary  because $\Gamma_0$ and $\Gamma_1$ are separately infrared divergent.   

We should add that in principle any consistent regularisation of the infrared divergences is acceptable. The main criterion for the choice of the infrared regulator will be determined by the precision of the terms remaining after  the cancellation of the infrared divergences  in a numerical simulation.

\section{Summary and Prospects}\label{sec:concs}

Lattice calculations of some hadronic quantities are already approaching (or even reaching) $O(1\%)$ precision and we can confidently expect that the uncertainties will continue to be reduced in future simulations. At this level of precision, isospin-breaking effects, including electromagnetic corrections, must 
be included in the determination of the relevant physical quantities. In this paper we present, for the first time, a method to compute electromagnetic effects in hadronic processes. For these quantities the presence of infrared divergences in the intermediate stages of the calculation makes the procedure much more complicated  than is the case for the hadronic spectrum, for which calculations in several  different approaches~\cite{Borsanyi:2014jba,deDivitiis:2013xla,Basak:2014vca,Ishikawa:2012ix,Aoki:2012st,Blum:2010ym} 
already exist. In order to obtain physical decay widths (or cross sections) diagrams containing virtual photons must be combined with those corresponding to the emission of real photons. Only in this way are the infrared divergences cancelled. We stress that it is not sufficient simply to add the electromagnetic interaction to the quark action  because, for any given process, the contributions corresponding to different numbers of real photons must be evaluated separately.

We have discussed in detail a specific case, namely the $O(\alpha)$ radiative corrections to the leptonic decay of charged pseudoscalar mesons. The method can however, be extended to many other processes, for example to semileptonic decays. The condition for the applicability of our strategy is that there is a mass gap between the decaying particle and the intermediate states generated by the emission of the photon, so that all of these states have higher energies than the mass of the initial hadron (in the rest frame of the initial hadron). 

In the present paper, we  have limited the discussion to real photons with  energies which are much smaller than the QCD scale $\Lambda_{\mathrm{QCD}}$.  This is not a limitation of our method and in the future one can envisage numerical simulations of contributions to the inclusive width from the emission of real photons with energies which do resolve the structure of the initial hadron. Such calculations can be performed in Euclidean space under the same conditions as above, i.e. providing that there is a mass gap. 

In the calculation of electromagnetic corrections a general issue concerns finite-size effects.  In this respect, our method reduces to the calculation of infrared-finite, gauge-invariant quantities for which we expect the finite-size corrections to be comparable to those encountered in the computation of  the spectrum.  This expectation will be checked in forthcoming numerical studies and studied theoretically in chiral perturbation theory.  Indeed an analytical calculation of the finite-volume effects requires a detailed analysis of the form factors parametrising the structure dependent contributions (see. Eq.\,(\ref{eq:hsd})).

Although the implementation of our method is challenging, it is within reach of present lattice technology particularly as the relative precision necessary to make the results phenomenologically interesting is not exceedingly high. Since the effects we are calculating are, in general, of $O(1\%)$, calculating the electromagnetic corrections to a precision of 20\% or so would already be more than sufficient. As the techniques improve and computational resources increase, the determination of both the QCD and QED effects will become even more precise. We now look forward to implementing the method described in this paper in an actual numerical simulation.

\section*{Acknowledgements}

We are particularly grateful to W. Marciano, for helpful correspondence  on the renormalisation of the electromagnetic corrections, to M. Sozzi for advice on experimental photon resolutions and to F. Sanfillipo and  G.C. Rossi for discussions.  Work partially supported by the ERC-2010 DaMESyFla Grant Agreement Number: 267985,  by the MIUR (Italy) under a contract  PRIN10 and by STFC Grants ST/J000396/1 and ST/L000296/1. 

\appendix

\section{Matching between Lattice and W-regularisation}\label{sec:Wilsonmatching}

In this appendix we briefly describe the matching between the lattice and W regularisations, in perturbative QED for the complete basis of four-fermion operators 
\be
O_{XY} = (\bar d\, \Gamma_X\, u) \, (\bar\nu_\ell\, \Gamma_Y\, \ell) \equiv \Gamma_X \otimes \Gamma_Y\,,
\ee
where $\Gamma_{X,Y}$ are Dirac matrices.
We consider the following basis of five four-fermion operators given in Eq.\,(\ref{eq:5ops}):
\be
\begin{array}{ll}
O_1 = \gamma^\mu (1-\gamma^5) \otimes \gamma_\mu (1-\gamma^5)\,, \qquad  & O_2 = \gamma^\mu (1+\gamma^5) \otimes \gamma_\mu (1-\gamma^5) \,,\\
O_3 = (1-\gamma^5) \otimes (1+\gamma^5)\,, \qquad  & O_4  = (1+\gamma^5) \otimes (1+\gamma^5) \,,\\
&\hspace{-1in}O_5 = \sigma^{\mu \,\nu}(1+\gamma^5) \otimes \sigma_{\mu \,\nu}(1+\gamma^5)\,.
\label{eq:O1O5}
\end{array}
\ee

The complete basis is made up of ten operators. The five additional operators are obtained from $O_1$\,-\,$O_5$  by the exchange $(1-\gamma^5) \leftrightarrow (1+\gamma^5)$.
Since the neutrino is electrically neutral its chirality is conserved and the operators $O_1$\,-\,$O_5$ do not mix under renormalisation with the remaining 5 operators and invariance under parity transformations ensures that the two $5 \times 5$ renormalisation matrices are equal. 
For this reason, in the following we focus the discussion on the five operators of Eq.~(\ref{eq:O1O5}). Moreover, the basis of operators in Eq.\,\ref{eq:O1O5} is the complete basis of operators for a left-handed neutrino.

With regularisations which respect chiral symmetry the four-fermion operator relevant for the leptonic weak decay, $O_1$, renormalizes multiplicatively. In this appendix we are using the lattice theory with Wilson fermions to illustrate the matching between the lattice and W-regularisations and the explicit 
breaking of chiral symmetry with this discretisation of QCD leads to the mixing of $O_1$ with the other four operators $O_2$\,-\,$O_5$. If instead of using Wilson fermions, we used a lattice formulation with good chiral properties, such as domain wall fermions, the corresponding discussion to the one presented below would be restricted to the single operator $O_1$ which transforms as the (8,1) representation under SU(3)$_\mathrm{L}\times$SU(3)$_\mathrm{R}$ chiral symmetry for the quarks. 

\begin{figure}[t]
\includegraphics[scale=1.00]{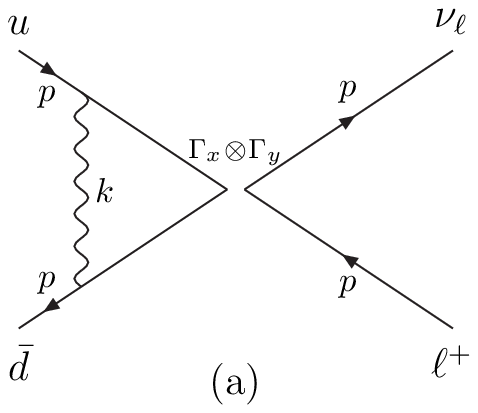}\qquad
\includegraphics[scale=1.00]{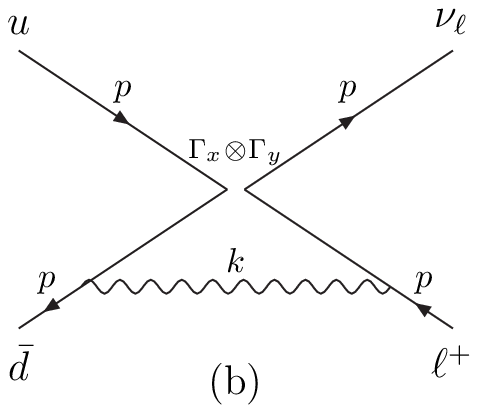}\qquad
\includegraphics[scale=1.00]{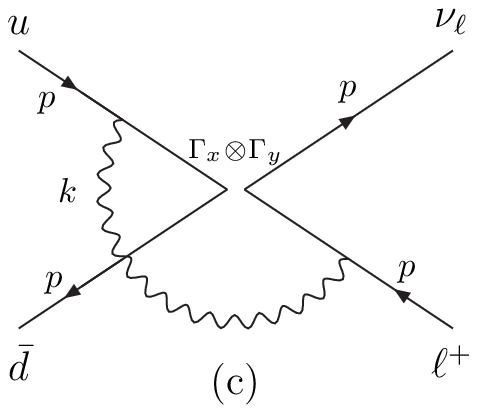}
\caption{One-loop Feynman diagrams computed for the renormalisation of the four-fermion operators $O_{XY} = (\bar d\, \Gamma_X\, u) \, (\bar\nu_\ell\, \Gamma_Y\, \ell) \equiv \Gamma_X \otimes \Gamma_Y$. \label{fig:diagrams}}
\end{figure}

We define $Z_{ij}(aM_W)$ to be the matrix which relates the operators $O_i$ (i=1\,-\,5) in the lattice and $W$-regularisations:  
\begin{equation}
O_i^W(M_W)=Z_{ij}(aM_W)\,O_j^\mathrm{latt}(a)\,.
\end{equation}
In order to perform the matching we adapt the RI-MOM renormalisation procedure developed for QCD~\cite{Martinelli:1994ty}, although, as described below, 
all the calculations here are performed in perturbation theory. Let $\Lambda^\mathrm{latt}_i$ and $\Lambda^W_i$ ($i$=1\,-\,5) be the amputated 4-quark Green function of the operator $O_i$ with the lattice and $W$ regularisations respectively, both with external momenta $p$ as illustrated in Fig.\,\ref{fig:diagrams}. We determine $Z$ by imposing that 
\begin{equation}\label{eq:rencond}
\left(Z_{u}^{-\frac12} \, Z_{d}^{-\frac12} \, Z_{\ell}^{-\frac12}\right)\,Z_{ik}\mathrm{Tr}\,(\Lambda^\mathrm{latt}_kP_j)=\mathrm{Tr}\,(\Lambda^\mathrm{W}_iP_j)\,.
\end{equation}
The projectors $P_j$  are defined by their action on the tree-level Green function $\Lambda_{i}^{(0)}$,
\be
\mbox{Tr}\left(\Lambda_{i}^{(0)}P_{j}\right)=\delta_{ij} \ ,
\ee
where the trace here and in Eq.~(\ref{eq:rencond}) is defined by $\mbox{Tr}\left(\Lambda_{i}P_{j}\right)=\mbox{Tr}\left(\Gamma^i_X\, P^j_X\, \Gamma^i_Y\, P^j_Y \right)$ for $O_{i}= \Gamma^i_X \otimes \Gamma^i_Y$ and $P_{j}=P^j_X \otimes P^j_Y$. $Z_{u,d,\ell}$ are the matching factors for the wave function renormalisation constants of the corresponding fermion fields, e.g. $u^W=Z_u^\frac12\,u^\mathrm{latt}$.

Consider the perturbative expansion of the amputated bare Green function in powers of the electromagnetic coupling in either the lattice or $W$-regularisations, 
\be
\Lambda_{i}=\Lambda_{i}^{(0)}+ \frac{\alpha}{4\pi}\, \Lambda_{i}^{(1)}+... \,.
\ee
In order to implement the matching conditions between the two regularisation schemes we require the quantities $\mbox{Tr}\left(\Lambda_{i}P_{j}\right)$ in both schemes. At one-loop order we write
\be
\label{eq:dynamical}
\mbox{Tr}\left(\Lambda_{i}^{\mathrm{latt}\,(1)}P_{j}\right)\equiv D_{ij} \qquad \mathrm{and} \qquad 
\mbox{Tr}\left(\mbox{\ensuremath{\Lambda}}_{i}^{W\,(1)}P_{j}\right)\equiv C_{ij} \,.
\ee
We represent the matching of the wave functions in the lattice and $W$-regularisations up to one-loop order by $Z_{q} = 1+(\alpha/4\pi)\,  Z_{q}^{(1)}+...$\,. Using Eq.\,(\ref{eq:rencond}), we see that  the matching matrix of the operators in Eq.\,(\ref{eq:O1O5}) at $O(\alpha)$ is given by
\be\label{eq:zij}
Z_{ij}^{(1)}=C_{ij}-D_{ij}+\dfrac{1}{2}\left(Z_{u}^{(1)}+Z_{d}^{(1)}+Z_{\ell}^{(1)}\right)\delta_{ij} \,.
\ee

We have presented the $\mathcal{O}(\alpha)$ contribution to the matching factor for the wave function of the charged lepton in Eq.\,(\ref{eq:zlmatching}) of Sec.\,\ref{sec:structure}:
\be
Z_{\ell}^{(1)}=-3/2-\log(a^2 M_W^2)-11.852\,,
\label{eq:Zmu}
\ee
$Z_u^{(1)}$ and $Z_d^{(1)}$ differ from $Z_\ell^{(1)}$ only by factors of $Q_u^2$ and $Q_d^2$, where $Q_f$ is the charge of the fermion $f$. We have verified with an explicit calculation that the contribution to the matching given in eq.(A9) is the same whether evaluated for an on-shell or an off-shell external lepton.

In order to evaluate the matrices $C_{ij}$ and $D_{ij}$ it is necessary to compute the Feynman diagrams shown in Fig.~\ref{fig:diagrams} in the two regularisation schemes.
All the external momenta are chosen to be equal to $p$ and all external particles are taken to be massless.
We deduce $D_{ij}$ from the results of the corresponding QCD calculation performed in~\cite{Constantinou:2010zs}. (Ref.~\cite{Constantinou:2010zs} includes a package containing an ASCII file, in order to make the results most easily accessible to the reader.) Diagrams 1, 2, 3 of Fig.~\ref{fig:diagrams} correspond to the diagrams $d_{5}$, $d_{6}$ and $d_{1}$ of \cite{Constantinou:2010zs}.
The expression for the lattice wave function renormalisation can be obtained from \cite{Constantinou:2009tr}. 

We now present results for the standard Wilson fermions and the ``na\"ive'' QED gauge action, for which the tree-level lattice photon propagator in the Feynman gauge is given in Eq.\,(\ref{eq:pp})\,. In infinite volume the sum over momenta in Eq.\,(\ref{eq:pp}) is replaced by the corresponding integral.
By combining the ingredients discussed above, we obtain the following result for the $O(\alpha)$ contribution to the renormalisation matrix $Z_{ij}$ of Eq.\,(\ref{eq:zij}):
\bea
& Z^{(1)}= \left(
\begin{array}{ccccc}
 2 L_W-15.539 & 0.536 & 1.607 & -3.214 & -0.804  \\
 0.536 & L_W-14.850 & -3.214 & 1.607 & -0.402  \\
 0.402 & -0.804 & -\frac{2}{3}L_W-13.702 & -1.071 & 0 \\
 -0.804 & 0.402 & -1.071 & -\frac{2}{3} L_W-13.702 & \frac{1}{12} L_W-0.057  \\
 -9.643 & -4.822 & 0 & 4 L_W-2.756 & \frac{20}{9} L_W-15.692 
\end{array}
\right)\!,  \nonumber \\ &
\label{eq:ZWilson}
\eea
where $L_W=\log(a^2 M_W^2)$.

The four-fermion operator relevant for the leptonic decay rate is $O_1$. From Eq.\,(\ref{eq:ZWilson}) we obtain the expression in Eq.\,(\ref{eq:matchingWilson}) for $O_1$ in the $W$-regularisation in terms of the bare lattice operators.

The result presented for $Z$ in Eq.\,(\ref{eq:ZWilson}) above is also valid if the twisted-mass (or Osterweilder-Seiler~\cite{Osterwalder:1977pc}) lattice regularisation is used for the fermions instead of the Wilson action. 
This statement follows from the observation that the twisted mass action, in the so called \emph{twisted basis}~\cite{Frezzotti:2003ni}, only differs from the Wilson action by the presence of $\gamma^5$ in the mass term. The two actions are therefore identical in the chiral limit and all renormalisation constants are equal for Wilson and twisted-mass fermions in the twisted basis in all mass-independent renormalisation schemes. The renormalisation constants for twisted-mass fermions in the physical basis  are obtained from those in the twisted basis through a simple twisted rotation~\cite{Frezzotti:2003ni}.

The lattice results in~\cite{Constantinou:2010zs,Constantinou:2009tr} are also given for a number of pure gauge actions including the tree-level Symanzik and Iwasaki actions.
For completeness we give below the results for the renormalisation matrix for these two choices of the gauge action:
\bea
& Z_{TS}^{(1)}= \left(
\begin{array}{ccccc}
 2 L_W-12.399 & 0.451 & 1.354 & -2.709 & -0.677 \\
 0.451 & L_W-11.866 & -2.709 & 1.354 & -0.339 \\
 0.339 & -0.677 & -\frac{2}{3} L_W-10.978 & -0.903 & 0 \\
 -0.677 & 0.339 & -0.903 & -\frac{2}{3} L_W-10.978 & \frac{1}{12} L_W-0.044 \\
 -8.127 & -4.063 & 0 & 4 L_W-2.132 & \frac{20}{9} L_W-12.518 
\end{array}
\right)\!,  \nonumber \\ &
\label{eq:ZS}
\eea
\bea
& Z_{Iw}^{(1)}= \left(
\begin{array}{ccccc}
2 L_W-11.732 & 0.323 & 0.969 & -1.938 & -0.485 \\
 0.323 & L_W-11.525 & -1.938 & 0.969 & -0.242 \\
 0.242 & -0.485 & -\frac{2}{3} L_W-11.181 & -0.646 & 0 \\
 -0.485 & 0.242 & -0.646 & -\frac{2}{3} L_W-11.181 &
  \frac{1}{12} L_W-0.017 \\
 -5.815 & -2.908 & 0 & 4 L_W-0.826 & \frac{20}{9} L_W-11.777 
\end{array}
\right)\!.  \nonumber \\ &
\label{eq:ZI}
\eea

\section{Structure dependent contributions to the real decay}\label{sec:sd}

In this appendix we estimate the size of the neglected structure-dependent contributions to the decay $P^+\to\ell^+\nu_\ell\gamma$
for light mesons, $P^+=\pi^+,K^+$. 
We base our estimates on the results of the phenomenological analyses performed in Refs.\,\cite{Bijnens:1992en,Bijnens:1994me,Cirigliano:2007xi} based on the use of chiral perturbation theory at $O(p^4)$. 
Although the relevant expressions have also been derived at $O(p^6)$~\cite{Ametller:1993hg,Geng:2003mt} (see also page 10 of \cite{Cirigliano:2011ny}), in that case there are too many unknown low-energy constants to be useful in making an estimate. As was done in the main body of the paper, for the general framework we give the explicit formulae for pion decays; the generalisation of the framework to kaons, and indeed also to $D$-mesons and $B$-mesons decays is straightforward. We then make the numerical estimates of the structure dependent effects for pions and kaons based on chiral perturbation theory. Finally we make some comments about structure dependent terms when $P^+$ is a heavy-light meson, $D^+$ or $B^+$.

The starting point of the analysis is the decomposition in terms of Lorenz invariant form factors of the hadronic matrix element (see also Eq.\,(\ref{eq:diagsef1}))
\begin{eqnarray}
H^{\mu\nu}(k,p_\pi)&=& \int d^4x\, e^{ikx}\, T\bra{0}j^\mu(x) J_W^\nu(0) \ket{\pi(p_\pi)} \;.
\end{eqnarray}
We follow the standard convention of separating the contribution corresponding to the approximation of a point-like pion (also frequently called \emph{inner bremsstrahlung}) $H^{\mu\nu}_\textrm{pt}$, from the structure dependent part $H^{\mu\nu}_\textrm{SD}$,
\begin{equation}\label{eq:hmunusep}
H^{\mu\nu}=H^{\mu\nu}_{\textrm{SD}} + H^{\mu\nu}_\textrm{pt} \;.
\end{equation}
$H^{\mu\nu}_\textrm{pt}$ is simply given by
\begin{equation}
H^{\mu\nu}_\textrm{pt} =
f_\pi\, \left[g^{\mu\nu}-\frac{(2p_\pi-k)^\mu(p_\pi-k)^\nu}{(p_\pi-k)^2-m_\pi^2} \right] \,.
\end{equation}
The structure dependent component can be parametrised by four independent invariant form factors which we define as
\begin{eqnarray}
H^{\mu\nu}_\textrm{SD} 
&=&
H_1\, \left[k^2g^{\mu\nu}-k^\mu k^\nu\right]+H_2\, \left\{\left[(k\cdot p_\pi-k^2)k^\mu-k^2(p_\pi-k)^\mu\right]\left(p_\pi-k\right)^\nu\right\}
\nonumber \\[0.15cm]
&&\hspace{0.5in}-
i\frac{F_V}{m_\pi}\, \epsilon^{\mu\nu\alpha\beta} k_\alpha {p_\pi}_\beta 
+
\frac{F_A}{m_\pi}\, \left[(k\cdot p_\pi -k^2)g^{\mu\nu}-(p_\pi-k)^\mu k^\nu \right] \;.
\label{eq:hsd}\end{eqnarray}
Note that the vector Ward Identity $k_\mu\, H^{\mu\nu} = f_\pi\, p_\pi^\nu$, derived in Ref.\,\cite{Bijnens:1992en},  is saturated by $H^{\mu\nu}_\textrm{pt}$
\begin{equation}
k_\mu\, H^{\mu\nu}_\textrm{pt} = f_\pi\, p_\pi^\nu \;,
\qquad
k_\mu\, H^{\mu\nu}_\textrm{SD} = 0 \;.
\end{equation}
As discussed in the main body of the paper,  $H_\textrm{pt}^{\mu\nu}$ also contains the infrared divergences which appear in the virtual- and real-photon contributions to the decay rate. 
These observations motivate the decomposition in Eq.\,(\ref{eq:hmunusep}).

In the calculation of the decay rate for $\pi^+\to\ell^+\nu_\ell\gamma$ the tensor $H^{\mu\nu}$ is contracted with the polarisation vector of the real photon. In physical gauges with 
$\varepsilon^\star \cdot k=0$ we define
\begin{equation}
H^\nu\equiv \varepsilon^\star_\mu H^{\mu\nu} \;,
\end{equation}
so that 
\begin{eqnarray}
H^\nu_{SD} 
&=&
-\varepsilon^\star_\mu
\left\{
i\frac{F_V}{m_\pi}\, \epsilon^{\mu\nu\alpha\beta} k_\alpha {p_\pi}_\beta 
-
\frac{F_A}{m_\pi}\, \left[(k\cdot p_\pi -k^2)g^{\mu\nu}-(p_\pi-k)^\mu k^\nu \right]
\right\} \;,
\end{eqnarray}
showing that the structure dependent part of the decay rate can be parametrized in terms of the two form factors $F_V$ and $F_A$.

Before performing the integrations over the three-body phase space, the differential decay rate can be expressed as a function of the two independent Dalitz variables ($p_\pi=p_\ell+p_\nu+k$)
\begin{eqnarray}
x_\ell= -\frac{(p_\pi-p_\ell)^2}{m_\pi^2}+1\;,
\qquad\textrm{\textrm{}}
x_\gamma= -\frac{(p_\pi-k)^2}{m_\pi^2}+1\;.
\end{eqnarray}
The decay rate as a function of the photon's energy in the pion's rest frame can be obtained by performing the integration over $x_\ell$ with the limits $x_\ell \in \left[x_\ell^{\textrm{min}},x_\ell^{\textrm{max}} \right]$ where
\begin{eqnarray}
x_\ell^{\textrm{min}} &=&  1-r_\gamma^2 - 
\frac{1-x_\gamma-r_\ell^2}{2(1-x_\gamma)}
\left[
x_\gamma-r_\gamma^2 + \sqrt{(x_\gamma+r_\gamma^2)^2-4r_\gamma^2}
\right] \; ,
\nonumber \\[0.15cm]
x_\ell^{\textrm{max}} &=&  1-r_\gamma^2 - 
\frac{1-x_\gamma-r_\ell^2}{2(1-x_\gamma)}
\left[x_\gamma-r_\gamma^2 - \sqrt{(x_\gamma+r_\gamma^2)^2-4r_\gamma^2}\right] \,,
\label{eq:xlfulllimits}
\end{eqnarray}
$r_\ell=m_\ell/m_\pi$ and $r_\gamma=m_\gamma/m_\pi$.
The total decay rate is obtained by performing the integral over $x_\gamma$ in the range $x_\gamma \in \left[x_\gamma^{\textrm{min}},x_\gamma^{\textrm{max}} \right]$ with
\begin{equation}
x_\gamma^{\textrm{min}} =  r_\gamma(2-r_\gamma) \;, 
\qquad
x_\gamma^{\textrm{max}} =  1-r_\ell^2 \;. 
\label{eq:xgfulllimits}
\end{equation}
The photon's mass $m_\gamma$ was introduced in the definition of $r_\gamma$ to regulate the infrared divergences in the point-like contribution. For the structure dependent contribution, which is infrared finite
we can set $m_\gamma\to 0$ and simplify the above expressions by making the replacements
\begin{eqnarray}
x_\ell^{\textrm{min}} &\mapsto&  (1-x_\gamma) + \frac{x_\gamma r_\ell^2}{(1-x_\gamma)} 
\; ,
\qquad
x_\ell^{\textrm{max}} \mapsto  1\; .
\nonumber \\
\nonumber \\
x_\gamma^{\textrm{min}} &\mapsto&  0 \;, 
\qquad
x_\gamma^{\textrm{max}} \mapsto  1-r_\ell^2 \;.
\label{eq:xlreduced}
\end{eqnarray}
The different contributions to the differential decay rate have been obtained in Ref.\,\cite{Bijnens:1992en}. Writing $\Gamma_1=
\Gamma_1^\textrm{pt}+\Gamma_1^\textrm{SD}+\Gamma_1^\textrm{INT}$, where $\Gamma_1^\textrm{INT}$ is the contribution to the decay rate coming from the interference between the point-like and the structure-dependent amplitudes, we confirm the following results:
\begin{eqnarray}
\frac{4\pi}{\alpha\, \Gamma_0^{\textrm{tree}}}
\frac{d^2\Gamma_1^\textrm{pt}}{dx_\gamma dx_\ell}
&=& 
\frac{2\, f_\textrm{pt}(x_\gamma,x_\ell)}{(1-r_\ell^2)^2}\;,
\nonumber \\[0.15cm]
\frac{4\pi}{\alpha\, \Gamma_0^{\textrm{tree}}}\frac{d^2\Gamma_1^\textrm{SD}}{dx_\gamma dx_\ell}
&=& 
\frac{m_\pi^2\left\{\left[F_V(x_\gamma)+F_A(x_\gamma)\right]^2f_\textrm{SD}^+(x_\gamma,x_\ell)+
\left[F_V(x_\gamma)-F_A(x_\gamma)\right]^2f_\textrm{SD}^-(x_\gamma,x_\ell)
\right\}}{2f_\pi^2\, r_\ell^2(1-r_\ell^2)^2}\,,
\nonumber \\[0.15cm]
\frac{4\pi}{
\alpha\, \Gamma_0^{\textrm{tree}}
}
\frac{d^2\Gamma_1^\textrm{INT}}{dx_\gamma dx_\ell}
&=& 
-\frac{2m_\pi\left\{\left[F_V(x_\gamma)+F_A(x_\gamma)\right]f_\textrm{INT}^+(x_\gamma,x_\ell)
+
\left[F_V(x_\gamma)-F_A(x_\gamma)\right]f_\textrm{INT}^-(x_\gamma,x_\ell)
\right\}
}{f_\pi\, (1-r_\ell^2)^2}\, 
.\nonumber
\\[0.01cm]
\label{eq:decayratessd}
\end{eqnarray}
The functions appearing in Eq.\,(\ref{eq:decayratessd}) are
\begin{eqnarray}
f_\textrm{pt}(x_\gamma,x_\ell)
&=&
\frac{1-x_\ell}{x_\gamma^2(x_\gamma+x_\ell-1)}
\left[
x_\gamma^2+2(1-x_\gamma)(1-r_\ell^2)-
\frac{2x_\gamma r_\ell^2(1-r_\ell^2)}{x_\gamma+x_\ell-1}
\right] \;,
\nonumber \\[0.25cm]
f_\textrm{SD}^+(x_\gamma,x_\ell)
&=&
(x_\gamma+x_\ell-1)\left[
(x_\gamma+x_\ell-1+r_\ell^2)(1-x_\gamma)-r_\ell^2
\right] \;,
\nonumber \\[0.25cm]
f_{SD}^-(x_\gamma,x_\ell)
&=&
-(1-x_\ell)\left[
(x_\ell-1+r_\ell^2)(1-x_\gamma)-r_\ell^2
\right] \;,
\\[0.25cm]
f_\textrm{INT}^+(x_\gamma,x_\ell)
&=&
-\frac{1-x_\ell}{x_\gamma(x_\gamma+x_\ell-1)}
\left[
(x_\gamma+x_\ell-1+r_\ell^2)(1-x_\gamma)-r_\ell^2
\right] \;,
\nonumber \\[0.25cm]
f_\textrm{INT}^-(x_\gamma,x_\ell)
&=&
\frac{1-x_\ell}{x_\gamma(x_\gamma+x_\ell-1)}
\left[
x_\gamma^2+(x_\gamma+x_\ell-1+r_\ell^2)(1-x_\gamma)-r_\ell^2
\right] \;.\nonumber
\end{eqnarray}
Whilst we confirm the results of Ref.\,\cite{Bijnens:1992en}, we note that we disagree with the sign of the interference term $d^2\Gamma_1^\textrm{INT}/dx_\gamma dx_\ell$ given in Refs.\,\cite{Agashe:2014kda,Becirevic:2009aq}.

The sum of Eqs.~(\ref{eq:pionpion}), (\ref{eq:leplep}) and (\ref{eq:rleplep}) of the main body of the paper can also be obtained by integrating the point-like contributions over $x_\ell$ with the limits given in Eq.\,(\ref{eq:xlfulllimits}) and over $x_\gamma$ in the range $[r_\gamma(2-r_\gamma),r_E]$.  It will be useful below to define the following quantities,
\begin{eqnarray}\label{eq:Q1Adef}
Q_1^A(x_\gamma) &=& 
\frac{4\pi}{\alpha\, \Gamma_0^{\textrm{tree}}}\, \frac{d\Gamma_1^A(x_\gamma)}{dx_\gamma}\;,
\hspace{1in}
A=\{\textrm{pt,SD,INT}\}\;,
\\[0.15cm]
R_1^A(\Delta E) &=& 
\frac{\Gamma_1^{A}(\Delta E)}
{\Gamma_0^{\alpha,\textrm{pt}}+\Gamma_0^{(d),\textrm{pt}}+\Gamma_1^\textrm{pt}(\Delta E)}\, 
\;,
\qquad
A=\{\textrm{SD,INT}\}\;,
\end{eqnarray}
where $\Delta E=r_E\, m_\pi/2$
and $\Gamma_0^{\alpha,\textrm{pt}}$ and $\Gamma_0^{\textrm{(d)},\textrm{pt}}$ have been defined in the main body of the paper (see Eq.\,(\ref{eq:gammadeltaf})). Notice that the quantity in the  denominator of $R_1^A(\Delta E)$ is infrared finite (although it does depend on $M_W$, the ultraviolet cutoff in the $W$-regularisation).

In the following we use phenomenological parametrisations of the form factors in order to estimate the size of the structure-dependent contributions to the decay rate $\Gamma_1$. For the case of light mesons, we can use the results of the calculations of refs.~\cite{Bijnens:1992en,Bijnens:1994me,Cirigliano:2007xi} (see also ref.~\cite{Cirigliano:2011ny}) based on chiral perturbation theory and approximate the form factors as constants. At $O(p^4)$ in chiral perturbation theory,
\begin{equation}
F_V=\frac{m_P}{4\pi^2f_\pi}\quad\textrm{and}\quad  F_A=\frac{8m_P}{f_\pi}\,(L_9^r+L_{10}^r)\,,
\end{equation}
where $P=\pi$ or $K$ and $L_9^r,\,L_{10}^r$ are Gasser-Leutwyler coefficients.
The numerical values of these constants have been taken from the review by M.Bychkov and G.D'Ambrosio in Ref.\,\cite{Agashe:2014kda}; the values of $F_V$ and $F_A$ are 0.0254 and 0.0119 for the pion and 0.096 and 0.042 for the Kaon (for the pion these values of the form factors, obtained from direct measurements, can be found in the supplement to \cite{Agashe:2014kda} found in ~\cite{S008}).
\begin{figure}[!pt]
\begin{center}
\includegraphics[width=0.45\textwidth]{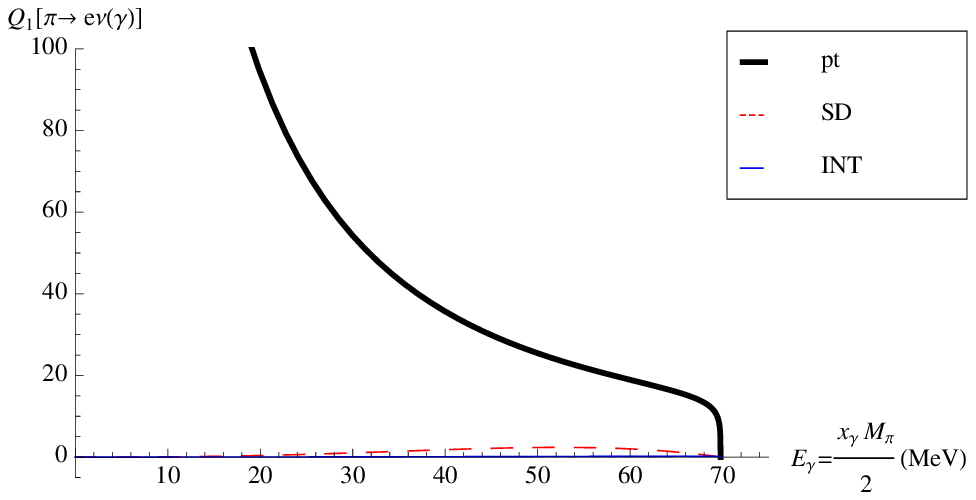}\qquad
\includegraphics[width=0.45\textwidth]{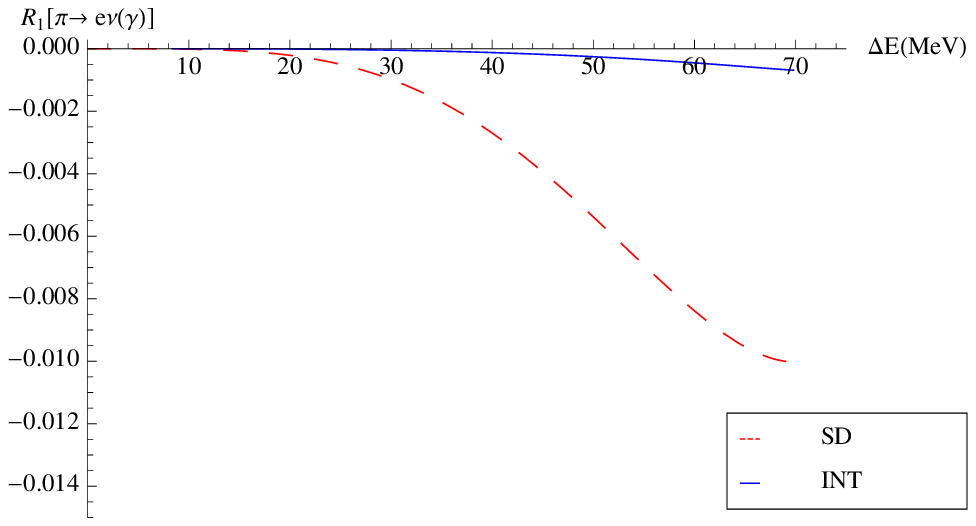}
\includegraphics[width=0.45\textwidth]{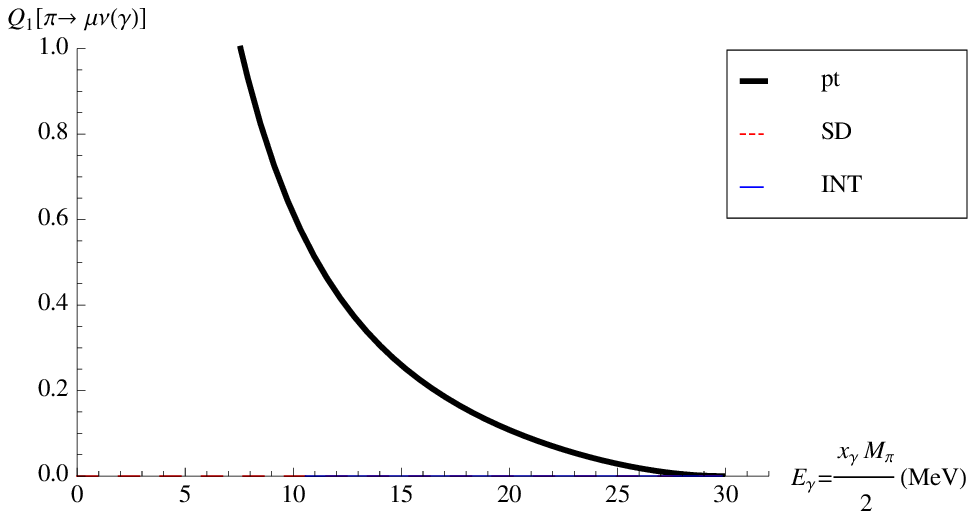}\qquad
\includegraphics[width=0.45\textwidth]{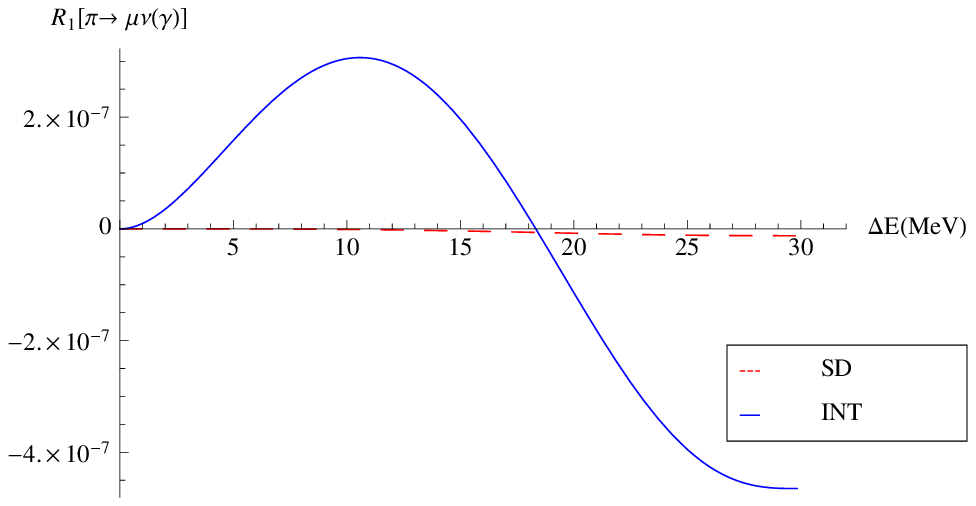}
\caption{
Point-like (pt), structure-dependent (SD) and interference (INT) contributions to the decay $\pi\rightarrow \ell\nu\gamma$. The first (second) row corresponds to $\ell=e$ ($\ell=\mu$). 
\label{fig:pion}}
\end{center}
\end{figure}
\begin{figure}[!h]
\begin{center}
\includegraphics[width=0.45\textwidth]{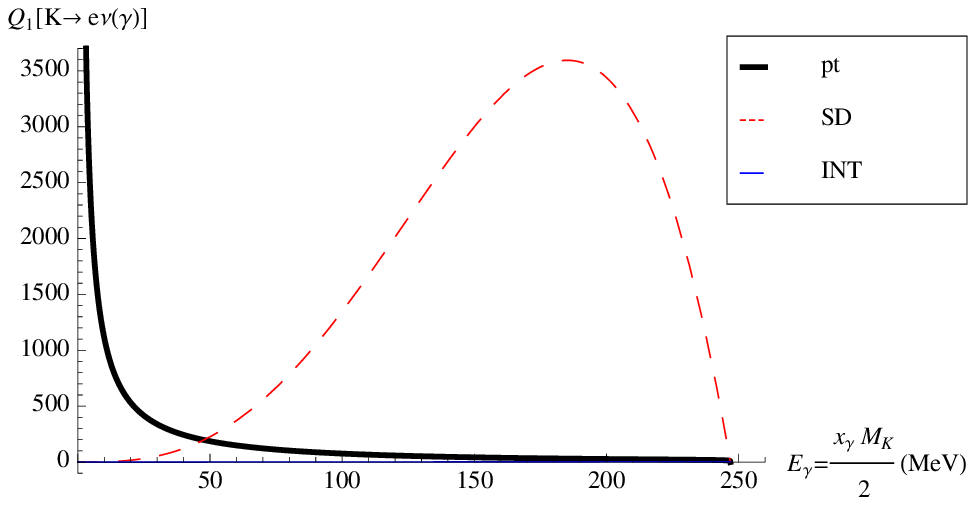}\qquad
\includegraphics[width=0.45\textwidth]{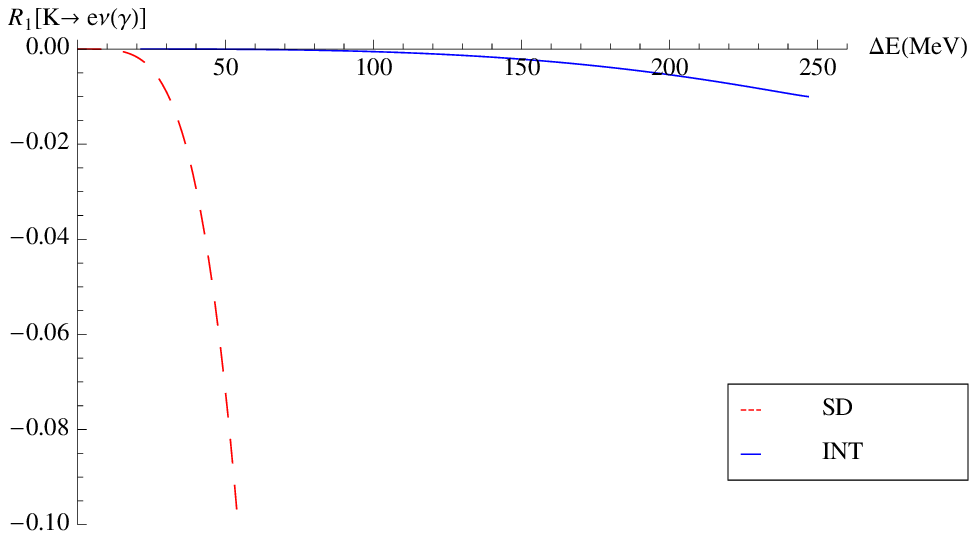}
\includegraphics[width=0.45\textwidth]{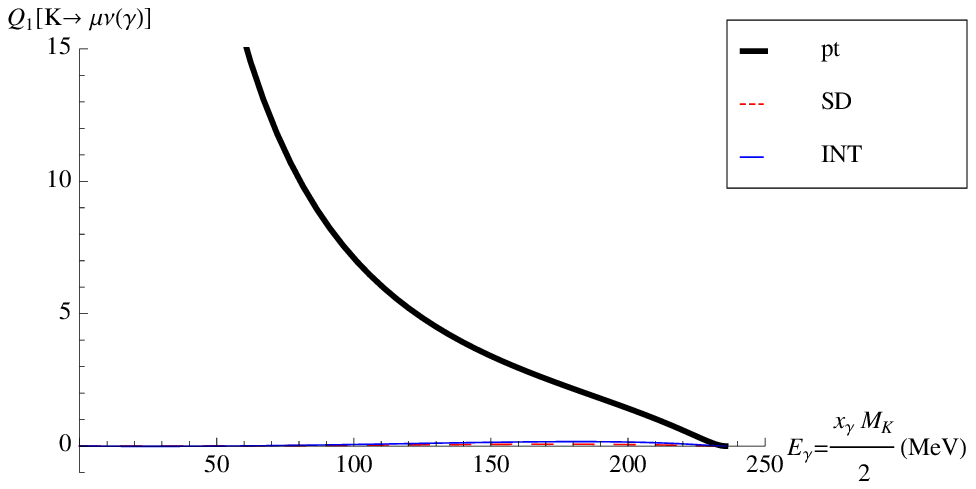}\qquad
\includegraphics[width=0.45\textwidth]{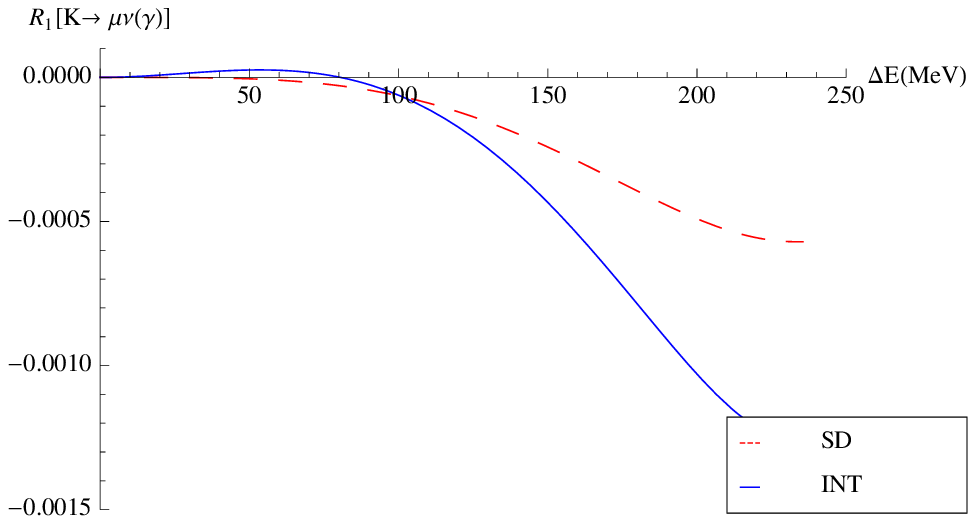}
\caption{
Point-like (pt), structure-dependent (SD) and interference (INT) contributions to the decay $K\rightarrow \ell\nu\gamma$. The first (second) row corresponds to $\ell=e$ ($\ell=\mu$). 
\label{fig:kaon}}
\end{center}
\end{figure}
In Figs.\,\ref{fig:pion} and \ref{fig:kaon} we compare the point-like, structure-dependent and interference contributions to the decays $\pi\rightarrow \ell\nu\gamma$ and $K\rightarrow \ell\nu\gamma$ respectively.  As can be seen, interference contributions are negligible in all the decays. The structure-dependent contributions can be sizeable because they are chirally enhanced with respect to the point-like contribution (notice the factor $1/r_\ell^2$ in the second equation in (\ref{eq:decayratessd})). From the phenomenological estimates of the form factors, this happens for the real decay $K\rightarrow e\nu_e\gamma$. On the other hand, for $E_\gamma<20$~MeV both structure dependent and interference contributions can be safely neglected with respect to the point-like contributions for all the decays of pions and the decay $K\to\mu\nu\gamma$. We learn from Refs.\,\cite{Ambrosino:2005fw,Ambrosino:2009aa} that a cutoff on the energy of the photon in the rest frame of the decaying particle of $O(20\,\textrm{MeV})$ is experimentally accessible.

\begin{figure}[!h]
\begin{center}
\includegraphics[width=0.325\textwidth]{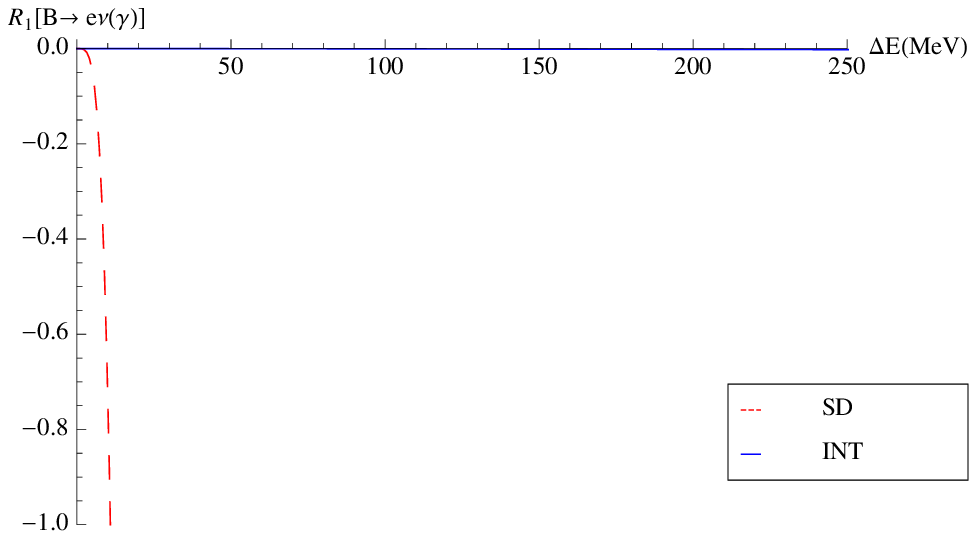}
\includegraphics[width=0.325\textwidth]{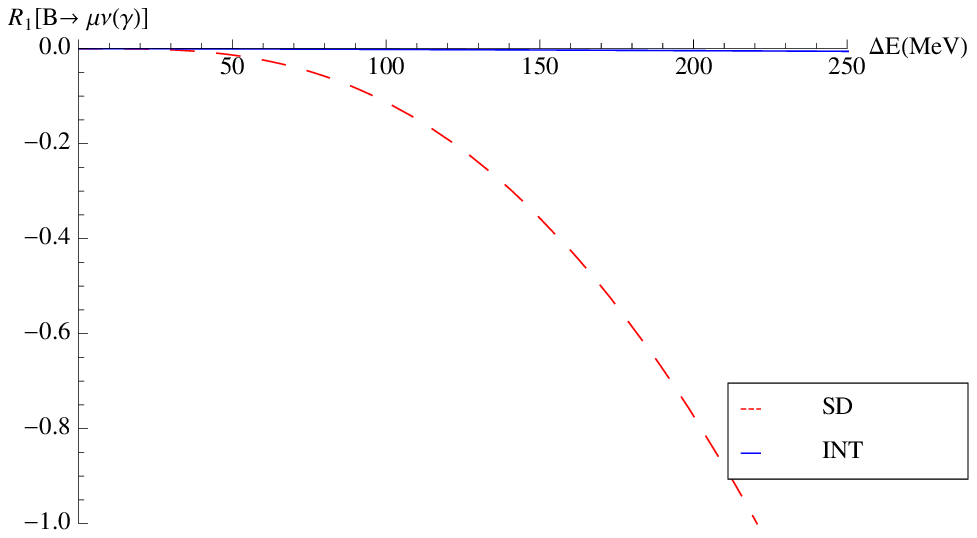}
\includegraphics[width=0.325\textwidth]{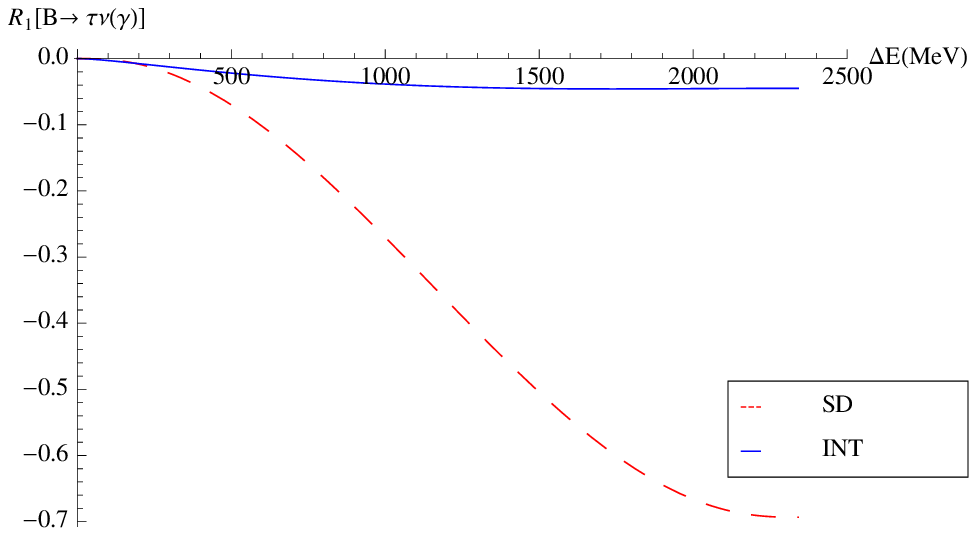}
\caption{
Structure-dependent (SD) and interference (INT) contributions to $R_1$ for the decays $B\rightarrow \ell\nu\gamma$. Going from left to right, the plots correspond to 
$\ell=e$, $\ell=\mu$ and $\ell=\tau$ respectively. \label{fig:bottom}}
\end{center}
\end{figure}

The application of chiral perturbation theory described above does not apply to the decays of $D$ and $B$ mesons and 
we believe that for these decays a lattice calculation of $F_{V,A}(x_\gamma)$ for a range of values of $x_\gamma$ will prove to be very useful as a check of the range of validity of the point-like approximation. As stressed in the main body of the paper, such a lattice calculation, starting from Euclidean correlators is indeed possible. A new feature in the case of $B$-decays in particular, one which is a consequence of the heavy-quark symmetry, is that the $B^\ast$ and $B$ are almost degenerate ($m_B^\ast-m_B\simeq 45$\,MeV). The radiation of a relatively soft photon can therefore cause the transition from a $B$-meson to an internal $B^\ast$ close to its mass-shell. Lattice calculations of the form factors would allow us to investigate the effect this small hyperfine splitting has on the size of the structure dependent terms as a function of $\Delta E$. 

In the absence of lattice calculations of the form factors, we note the phenomenological analysis of Ref.\,\cite{Becirevic:2009aq}, based on the extreme assumption of the single pole dominance, $B^\ast$ for $F_V$ and $B_1(5721)$ for $F_A$ (in reality many other virtual states contribute to the form factors):
\begin{eqnarray}
F_V(x_\gamma) \simeq \frac{C_V}{x_\gamma -1+m_{B^\star}^2/m_B^2}\,,\qquad F_A(x_\gamma) \simeq \frac{C_A}{x_\gamma -1+m_{B_1(5721)}^2/m_B^2}~,
\end{eqnarray}
with $C_V=0.24$ and $C_A=0.20$. The corresponding ratios $R_1$ are shown in Figure~\ref{fig:bottom}, from which it can be seen that under this assumption the structure-dependent contributions to $B\to e\nu_e\gamma$ for $E_\gamma\simeq20$\,MeV can be very large, but are small for $B\to \mu\nu_\mu\gamma$ and $B\to \tau\nu_\tau\gamma$ .

\end{document}